%% file: main.tex
\documentclass[prl,twocolumn,twoside,preprintnumbers,superscriptaddress,nofootinbib]{revtex4-2}

\usepackage{physics}
\usepackage{amsmath}
\usepackage{amssymb}
\usepackage{hyperref}
\usepackage{enumitem}
\usepackage{xspace}
\usepackage{slashed}
\usepackage{braket}
\usepackage{leftindex}
\usepackage{graphicx}  %
\usepackage{bm}  %                                          
\usepackage{amsmath}   %
\usepackage{graphics}%, slashed, braket}
\usepackage{color}
\usepackage{colortbl}
\usepackage{ulem}
\usepackage{hyperref}
\usepackage{xcolor}
\usepackage{comment}
\usepackage{titlesec}
\usepackage{enumitem}

\usepackage{graphicx}
\usepackage{array}
\usepackage{tabularray}
\usepackage[caption=false]{subfig}

\hypersetup{
    colorlinks=true,       % false: boxed links; true: colored links
    linkcolor=blue,          % color of internal links
    citecolor=blue,        % color of links to bibliography
    filecolor=blue,      % color of file links
    urlcolor=blue           % color of external links
}

\allowdisplaybreaks[1]

\titlespacing{\section}{0pt}{\parskip}{-\parskip}
\titlespacing{\subsection}{0pt}{\parskip}{-\parskip}
\titlespacing{\subsubsection}{0pt}{\parskip}{-\parskip}

\begin{document}

\begin{flushright}
LTH 1383
\end{flushright}
\vspace{-1cm}
\title{Muon $g$$-$$2$: blinding for data-driven hadronic vacuum polarization}

\author{Alexander~Keshavarzi}
\affiliation{Department of Physics and Astronomy, The University of Manchester, Manchester M13 9PL, U.K.}

\author{Daisuke~Nomura}
\affiliation{Department of Radiological Sciences, International University of Health and Welfare, Tochigi 324-8501, Japan}

\author{Thomas~Teubner}
\affiliation{Department of Mathematical Sciences, University of Liverpool, Liverpool L69 3BX, U.K.}

\author{Aidan~Wright}
\affiliation{Department of Mathematical Sciences, University of Liverpool, Liverpool L69 3BX, U.K.}

\begin{abstract}
The KNT(W) data-driven determinations of the hadronic vacuum polarization (HVP) are crucial inputs to previous and future Standard Model (SM) predictions of the muon's anomalous magnetic moment, $a_\mu$. With the muon $g$$-$$2$'s new physics case uncertain due to disagreeing HVP evaluations, new SM predictions and experimental measurements of $a_\mu$ expected soon, and a complete revamp of the KNTW analysis framework underway, this letter motivates and describes a blinding scheme for data-driven HVP determinations that has been implemented for future KNTW analyses. 

\end{abstract}

\maketitle

\input{sections/1_intro}
\input{sections/2_motivation}

\input{sections/3_procedure}

\input{sections/4_example}

\input{sections/5_conclusions}

{\bf Acknowledgements} -
Special thanks are extended to Mark~Lancaster for his input and for agreeing to be the external blinder for the new KNTW analysis. We would like to thank 
the Muon $g$$-$$2$ Theory Initiative and, in particular, the DHMZ group (Michel~Davier, 
Andreas~Hoecker, 
Bogdan~Malaescu,
Zhiqing~Zhang),
Antoine~G\'erardin
and 
Christoph~Lehner
for numerous useful discussions. AK is supported by The Royal Society (URF$\backslash$R1$\backslash$231503), STFC (Consolidated Grant
ST/S000925/) and the European Union’s Horizon 2020
research and innovation programme under the Marie
Sklodowska-Curie grant agreement No. 858199 (INTENSE). 
DN is supported by the Japan Society for the Promotion of Science under Grant Number 20K03960. TT is supported by STFC (Consolidated Grants ST/T000988/ and currently ST/X000699/). AW is supported by a PGR studentship jointly funded by STFC and the Leverhulme Trust under LIP-2021-01. 

\bibliographystyle{apsrev4-1.bst}
\bibliography{bib/1_intro,bib/2_motivation,bib/3_procedure}

\end{document}

%% file: sections/1_intro.tex
{\bf Introduction} - 
The anomalous magnetic moment of the muon,  $a_\mu$, and its potential for discovering new physics stand at a crossroads. The accuracy and precision of the Standard Model (SM) prediction, $a_\mu^{\rm SM}$~\cite{Aoyama:2020ynm,aoyama:2012wk,Aoyama:2019ryr,czarnecki:2002nt,gnendiger:2013pva,davier:2017zfy,keshavarzi:2018mgv,colangelo:2018mtw,hoferichter:2019mqg,davier:2019can,keshavarzi:2019abf,kurz:2014wya,melnikov:2003xd,masjuan:2017tvw,Colangelo:2017fiz,hoferichter:2018kwz,gerardin:2019vio,bijnens:2019ghy,colangelo:2019uex,Blum:2019ugy,colangelo:2014qya}, relies on resolving significant tensions in evaluations of the hadronic vacuum polarization (HVP) contributions, $a_\mu^{\rm HVP}$. Data-driven evaluations of the HVP using $e^+e^-\to{\rm hadrons}$ cross section data as input~\cite{Brodsky:1967sr,Lautrup:1968tdb,Krause:1996rf,kurz:2014wya,Jegerlehner:2017gek,Jegerlehner:2015stw,Jegerlehner:2017lbd,Jegerlehner:2017zsb,Jegerlehner:2018gjd,Eidelman:1995ny,Benayoun:2007cu,Benayoun:2012etq,Benayoun:2012wc,Benayoun:2015gxa,Benayoun:2019zwh,Benayoun:2019zwh,Davier:2010nc,Davier:2010nc,davier:2017zfy,Hagiwara:2003da,Hagiwara:2006jt,Hagiwara:2011af,keshavarzi:2018mgv,colangelo:2018mtw,hoferichter:2019mqg,davier:2019can,keshavarzi:2019abf,kurz:2014wya,melnikov:2003xd,masjuan:2017tvw,Colangelo:2017fiz,hoferichter:2018kwz} result in a value for $a_\mu^{\rm SM}$ that is $\sim5\sigma$ below the most recent experimental measurement from the Muon $g$$-$$2$ Experiment at Fermilab, $a_\mu^{\rm exp}$~\cite{Muong-2:2023cdq,Muong-2:2024hpx}. With an unprecedented 200 parts-per-billion (ppb) precision~\cite{Muong-2:2023cdq,Muong-2:2024hpx}, confirmation of previous measurements~\cite{Muong-2:2021ojo,Muong-2:2021xzz,Muong-2:2021ovs,Muong-2:2021vma,Muong-2:2006rrc,Muong-2:2002wip,Muong-2:2004fok}, and final results (expected in 2025) projected to improve the experimental precision by another factor of two, the measurements of $a_\mu$ appear to be on solid ground.\footnote{Alternative future measurements of $a_\mu$ are also planned at J-PARC~\cite{Abe:2019thb} and PSI~\cite{Adelmann:2021udj}.} However, high-precision lattice QCD calculations (incorporating QED corrections)~\cite{Aoyama:2020ynm,Budapest-Marseille-Wuppertal:2017okr,RBC:2018dos,Giusti:2019xct,FermilabLattice:2019ugu,Gerardin:2019rua,chakraborty:2017tqp,blum:2018mom,shintani:2019wai,Aubin:2019usy,giusti:2019hkz,Borsanyi:2020mff,Boccaletti:2024guq,RBC:2018dos,Lehner:2020crt,ExtendedTwistedMass:2022jpw,RBC:2023pvn,Kuberski:2024bcj,FermilabLattice:2022izv} and the most recent experimental measurement of the dominant $e^+e^-\to\pi^+\pi^-$ cross section from the CMD-3 experiment~\cite{CMD-3:2023alj,CMD-3:2023rfe} result in independent, but consistent values for $a_\mu^{\rm HVP}$ that are $>4\sigma$ larger than previous data-driven evaluations. They therefore generate values for $a_\mu^{\rm SM}$ that are consistent with $a_\mu^{\rm exp}$ and support a no-new-physics scenario in the muon $g$$-$$2$, whilst leaving an unexplained discrepancy with the vast catalogue of previously measured hadronic cross section data.

The KNT~\cite{keshavarzi:2018mgv,keshavarzi:2019abf} (now KNTW) data-driven determinations of $a_\mu^{\rm HVP}$ are crucial inputs to previous and future community-approved predictions for $a_\mu^{\rm SM}$ from the Muon $g$$-$$2$ Theory Initiative~\cite{Aoyama:2020ynm}. With multiple, independent lattice QCD evaluations of $a_\mu^{\rm HVP}$ becoming significantly competitive only in recent years, it was one of only a few data-driven HVP evaluations~\cite{colangelo:2018mtw,hoferichter:2019mqg,davier:2019can,keshavarzi:2019abf} which exclusively formed the value for $a_\mu^{\rm HVP}$ used in the SM prediction that exhibits the $\sim 5\sigma$ discrepancy with $a_\mu^{\rm exp}$~\cite{Aoyama:2020ynm,aoyama:2012wk,Aoyama:2019ryr,czarnecki:2002nt,gnendiger:2013pva,davier:2017zfy,keshavarzi:2018mgv,colangelo:2018mtw,hoferichter:2019mqg,davier:2019can,keshavarzi:2019abf,kurz:2014wya,melnikov:2003xd,masjuan:2017tvw,Colangelo:2017fiz,hoferichter:2018kwz,gerardin:2019vio,bijnens:2019ghy,colangelo:2019uex,Blum:2019ugy,colangelo:2014qya}. Future SM predictions are expected to incorporate both lattice QCD and updated data-driven evaluations, with KNTW being a key input to the latter. An alternative approach to determine $a_\mu^{\rm HVP}$ by experimentally measuring the spacelike vacuum polarization is under preparation at the MUonE Experiment~\cite{CarloniCalame:2015obs,Abbiendi:2016xup,Abbiendi:2677471}. 

The KNTW procedure for evaluating the total hadronic cross section and $a_\mu^{\rm HVP}$ (plus other precision observables which depend on hadronic effects) is undergoing a major overhaul and modernization of the analysis framework. The aim of this revamp is to make use of sophisticated analysis tools, perform new evaluations of various contributions, incorporate handles in the analysis structure that result in flexible and robust ways to test various systematic effects, improve determinations of corresponding systematic uncertainties and ultimately produce a new state-of-the-art in the determination of these quantities. These changes will be described in detail in the next full KNTW update.

Such future data-driven evaluations of $a_\mu^{\rm HVP}$ depend largely on new experimentally measured hadronic cross section data, particularly for the $\pi^+\pi^-$ final state. These require increased precision and a more robust understanding of higher-order radiative corrections, which are currently being studied in detail within the STRONG-2020 program~\cite{Strong2020} and The RadioMonteCarlow~2 Effort~\cite{RadioMonteCarlow} (see also e.g.~\cite{BaBar:2023xiy,Davier:2023fpl}). Whilst a discussion of these improvements is outside the scope of this letter, such future results have been announced from the BaBar~\cite{newBaBar}, Belle II~\cite{newBelleII}, BESIII~\cite{newBESIII}, CMD-3~\cite{newCMD3}, KLOE~\cite{newKLOE} and SND~\cite{newSND} experiments within the next few years. These new measurements could either fundamentally adjust the previous data-driven evaluations of $a_\mu^{\rm HVP}$ to bring them more in line with e.g.~the recent CMD-3 $\pi^+\pi^-$ measurement or make the current tensions even worse if new measurements confirm lower cross section values with increased precision. 

Importantly, and as will be discussed in the next section, analysis choices in how to use these data can produce significantly different results. With this being the case, the future of $a_\mu^{\rm HVP}$ and $a_\mu^{\rm SM}$ being so uncertain, and the crossroads in the current tensions ultimately suggesting either a discovery of new physics or a multi-method confirmation of the SM, analysis blinding for data-driven determinations of the HVP is now paramount. This is compounded by the fact that all other critical inputs: results from the Muon $g$$-$$2$ Experiment at Fermilab, lattice QCD calculations, and future $e^+e^-\to{\rm hadrons}$ cross section measurements, are blind analyses. As such, this letter motivates and describes the first blinding scheme for data-driven HVP determinations that has been implemented for future KNTW analyses.

% The KNTW evaluations of the total hadronic cross section, $a_\mu^{\rm HVP}$ and other precision observables dependent upon hadronic effects are based upon an analysis framework that combines the available data for  
% The continuing philosophy of the KNTW determinations is to provide maximally model-independent evaluations of the total hadronic cross section and $a_\mu^{\rm HVP}$, relying fully on the measured $e^+e^-\to{\rm hadrons}$ cross data to characterize the hadronic processes and avoiding alternative theoretical descriptions wherever possible.

%% file: sections/2_motivation.tex
{\bf Blinding motivation} - 
In a given data-driven analysis, $e^+e^-\to{\rm hadrons}$ cross section data from different experiments are combined in a statistically robust procedure. 
% The available data are measurements of either individual exclusive hadronic final states (certainly below $\sim$2 GeV where the hadronic physics is fully non-perturbative) or inclusive (all-hadrons) data. 
%Each distinct hadronic final state for which data can be combined is known as a channel. 
Many measurements can exist for each distinct hadronic final state (channel), each with different features: energy range, measurement technique, luminosity normalization, energy binning in $\sqrt{s}$, treatment of radiative corrections, prescription for provided experimental uncertainties, etc. The combination procedure then largely derives from four stages: (1) ensuring different data are consistent before combining, (2) defining a new energy binning onto which the input data will be combined, (3) combining the data in a fit procedure that is weighted in some form by the experimental uncertainties, and (4) applying additional systematic uncertainties arising from the combination procedure to the combined cross section. Derived quantities such as $a_\mu^{\rm HVP}$ are calculated from the combined data. Statistical comparisons can then be performed between the data combination and the input datasets for the cross section values and any calculated observables.

All four stages depend on analysis choices. In (1), for example, combining the data for the extraction of $a_\mu^{\rm HVP}$ necessitates the cross section data to be undressed of vacuum polarization effects and to include final state radiation, requiring the analyzer to define a procedure to ensure corrections for either or both are applied to any data that are not in that form. In stage (2), analyzer freedom to choose the target bin centers in $\sqrt{s}$, the bin width, and overall number of bins can lead to significantly different results, particularly in important hadronic resonance regions. For (3), the degree to which an analysis chooses to weight the fit/combination procedure by the experimental uncertainties, particularly regarding weighting with correlated uncertainties, can vastly change the influence of different data sets. Correspondingly, the resulting systematic uncertainties in (4) can depend on the choices in the earlier stages.

Together, the choices in these stages can lead to different results for the mean values and uncertainties for both the combined cross section and any derived quantity like $a_\mu^{\rm HVP}$. The combination procedure used in the previous KNT analyses are described in~\cite{keshavarzi:2018mgv,keshavarzi:2019abf} (and earlier in~\cite{Hagiwara:2003da,Hagiwara:2006jt,Hagiwara:2011af}). In general, however, none of the data-driven evaluations of the HVP (e.g.~\cite{Benayoun:2019zwh,colangelo:2018mtw,hoferichter:2019mqg,davier:2019can,keshavarzi:2019abf}) can avoid making or already having made these choices and correspondingly different results are observed between different evaluations. Such differences were explored in detail in~\cite{keshavarzi:2018mgv,Aoyama:2020ynm}. A prominent example is stage (3) for the dominant $\pi^+\pi^-$ channel, where it has been observed that choosing to maximally weight the fitted cross section by the correlated uncertainties favors the three high-precision, highly-correlated, lower-valued measurements from the KLOE experiment~\cite{KLOE:2008fmq,KLOE:2010qei,KLOE:2012anl,KLOE-2:2017fda} leading to an overall lower value of $a_\mu^{\pi^+\pi^-}$, whilst choosing minimal weighting of correlated uncertainties favors the single high-precision, narrowly-binned, but higher-valued measurement from the BaBar experiment~\cite{BaBar:2009wpw,BaBar:2012bdw}. This effect is the main reason for the two global, data-driven HVP analyses that featured in~\cite{Aoyama:2020ynm}: KNT19~\cite{keshavarzi:2019abf} and DHMZ19~\cite{davier:2019can}, yielding mean values for $a_\mu^{\pi^+\pi^-}$ that differ at the level of the final uncertainty (see Section 2.3.5 in~\cite{Aoyama:2020ynm}).

The future presents the potential for even larger differences. Already the CMD-3 $\pi^+\pi^-$ data~\cite{CMD-3:2023alj,CMD-3:2023rfe} is several standard deviations higher than all other $\pi^+\pi^-$ data and all new data-driven evaluations will have to include these data in their combinations. As mentioned previously, new higher-precision data could lead to even more exaggerated tensions, particularly for $\pi^+\pi^-$ where they could favor either the CMD-3 result or the previous data. There is also potential for some future data-driven analyses to resurrect the use of hadronic $\tau$-decay data~\cite{Davier:2023fpl,Masjuan:2023qsp} in their combinations and evaluations of $a_\mu^{\rm HVP}$. These historically exhibit higher results for $a_\mu^{\rm HVP}$ that are more in-line with the CMD-3 data and lattice QCD evaluations but have been avoided in recent years due to an incomplete understanding of the necessary isospin breaking corrections~\cite{Aoyama:2020ynm}.\footnote{Efforts are ongoing to calculate these isospin breaking corrections using lattice QCD (see e.g.~\cite{Bruno:2018ono}) which could make the hadronic $\tau$-decay data competitive within the next few years.}

In general, all the above has the potential for analysis bias. In the face of significant tensions from current and possibly future data, this applies to any previously decided analysis choices and any new ones, as either may consciously or unconsciously bias a data-driven analysis towards one result or another. Given that the new physics case in $a_\mu$ rests upon resolving the tensions in $a_\mu^{\rm HVP}$ which are currently reflected in the $e^+e^-\to{\rm hadrons}$ data, the future of its data-driven determinations requires unbiased results achieved through fully blind analyses that have re-evaluated all choices and any corresponding systematic uncertainties.

%% file: sections/3_procedure.tex
{\bf Blinding procedure} - 
Blinding an analysis of already measured data presents individual challenges. Retrospectively blinding publicly available experimental data and corresponding results is impossible as many experimental analyses make comparisons with previous data and calculate quantities like $a_\mu^{\rm HVP}$. The aim is therefore to implement a robust procedure that meaningfully blinds the outputs of any data combinations without introducing new biases to or interfering with the data combination methodology. Blinding by adjusting the input data that enter a combination is consequently avoided.\footnote{The ability to scrutinize input datasets when not combining or comparing to a combination is retained.} 

Instead, offsets are applied to the combined cross section to blind all visual and other outputs of the analysis, namely plots of the combined data and any values derived from them (e.g. $a_\mu^{\rm HVP}$). Importantly, no combined data are saved without blinding offsets applied. Defining the experimentally measured hadronic cross section as $\sigma_{\rm had}(s) \equiv e^+e^-\to{\rm hadrons}$, the offsets adjust the cross section for a given hadronic channel $i$ as
\begin{equation}\label{eq:sigma_blind}
    \sigma^{\rm blind}_{{\rm had},i}(s) = a_i\,b_i\,(s+s_{0,i})^{c_i} \, \sigma_{{\rm had},i}(s) \, .
\end{equation}
A different set of offsets for each channel $i$ ensures that a change in one channel cannot be disentangled by knowledge of a change in another. The subscript $i$ is suppressed in the following for simplicity of notation. The offsets $a$, $b$, $c$, and $s_0$ apply an amalgamation of the following: change in overall sign ($a$), an energy-independent multiplicative scale factor to conceal changes in size ($b$), power adjustments to the energy-dependence ($c$) and additive adjustments to the energy-dependence ($s_0$). In a given channel, the data used to calculate derived values and those entering plots are blinded with different offsets.
%\footnote{In any comparison of a resulting data combination with its input datasets, the input datasets are adjusted by the same blinding scheme to determine a comparable derived quantity or plot (see e.g. Fig.~\ref{fig:pipirho_plots}). The unknown energy-dependent adjustments ensure that no comparison of a combination with input datasets can result in full knowledge of the influence of that dataset on a combination or consequent extraction of the blinding offsets. No input datasets are saved with blinding offsets applied.} 
This provides additional relative blinding between plots of the data and any derived quantities. 

\begin{figure*}[!t]
  \centering
  \subfloat[The $\rho$ resonance of the $\pi^+\pi^-$ channel.]{%
  \label{fig:pipirho_plots}
    \includegraphics[width=0.43\textwidth]{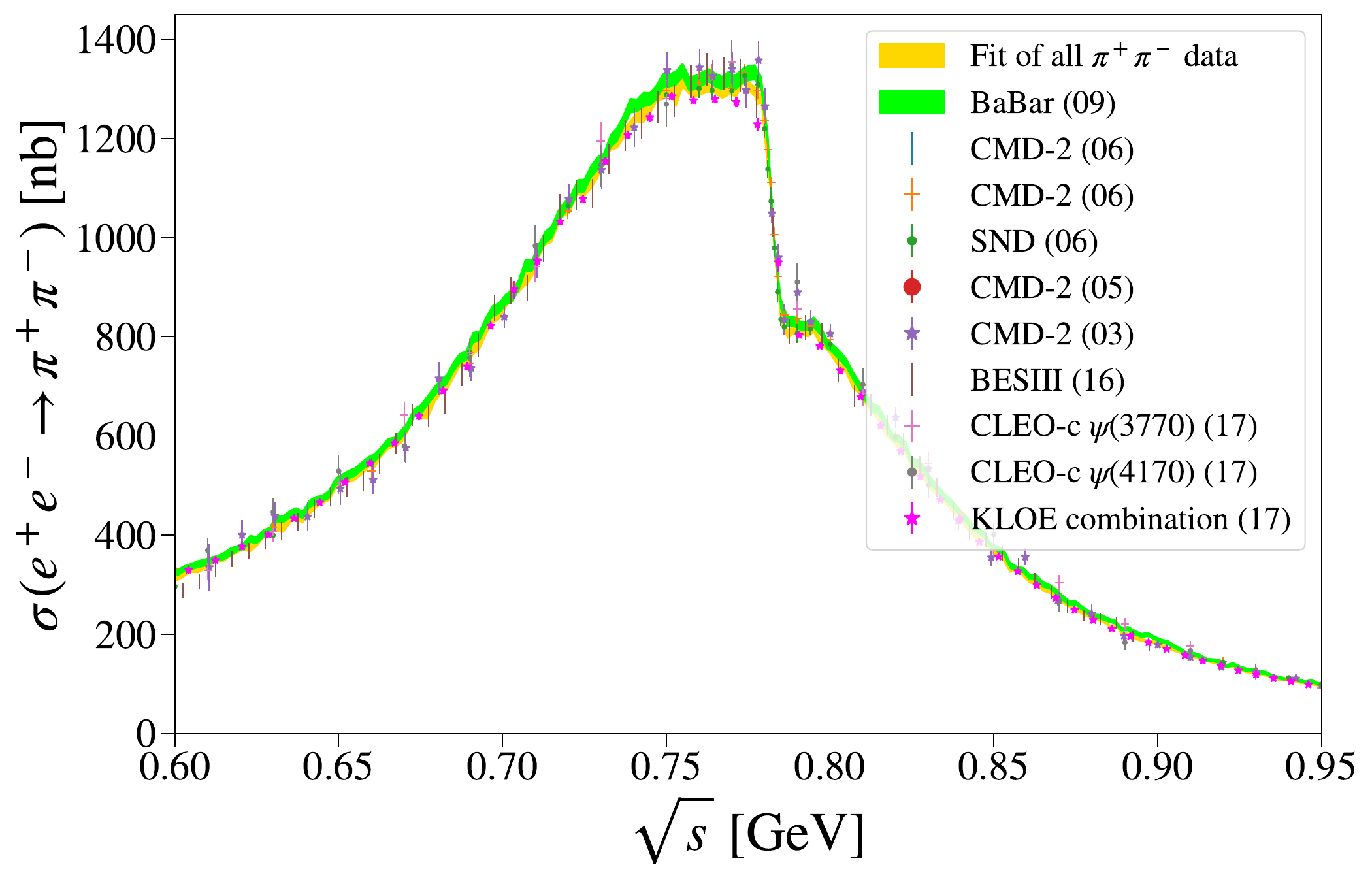}
    \hspace{0.4cm}
    \includegraphics[width=0.43\textwidth]{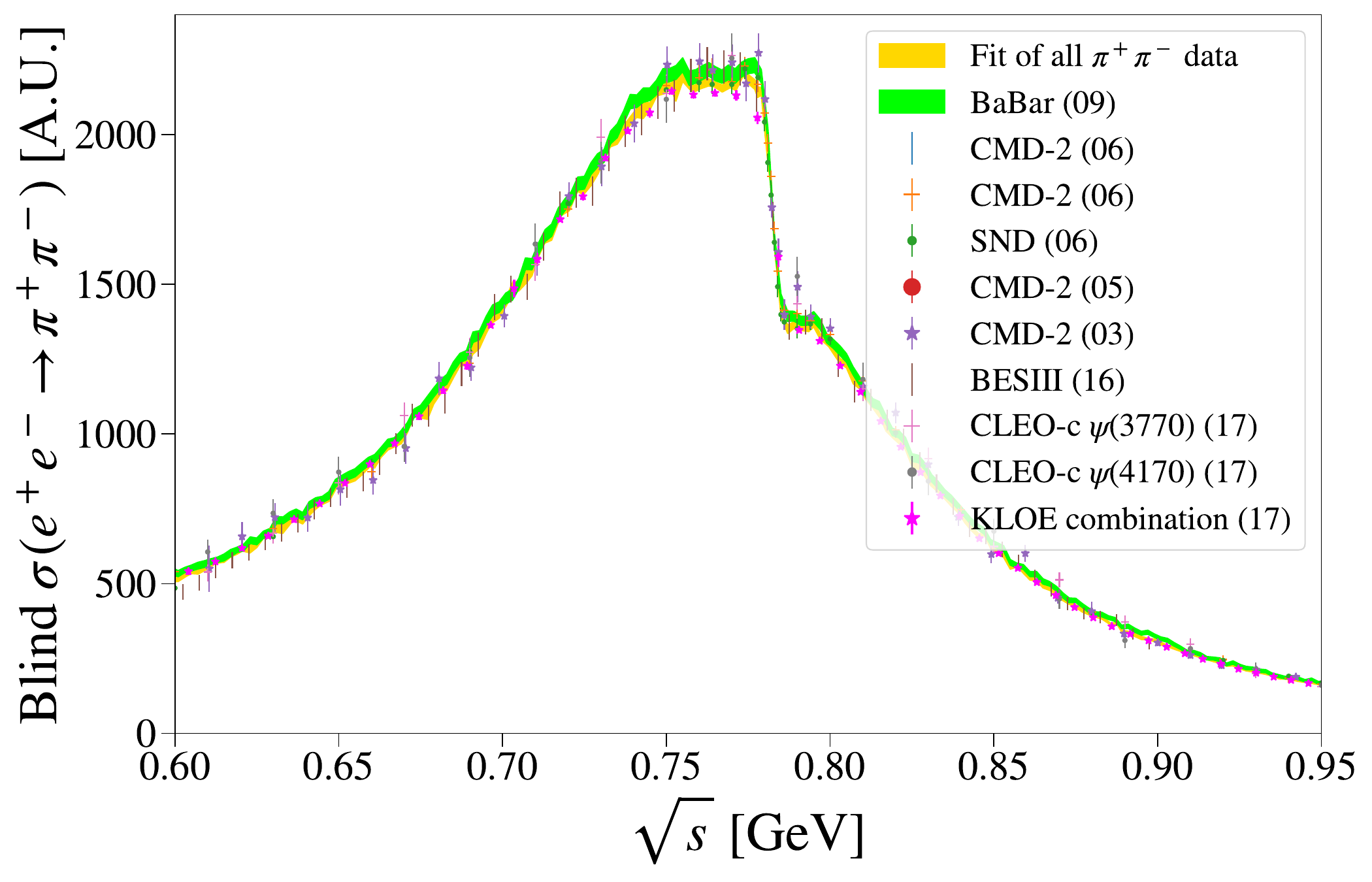}
    }\hfill 
  \subfloat[Comparison of values $a_\mu^{\pi^+\pi^-}$ in the range $0.6 \leq \sqrt{s} \leq 0.9$ GeV.]{% 
    \label{fig:pipirho_ints}
    \includegraphics[width=0.43\textwidth]{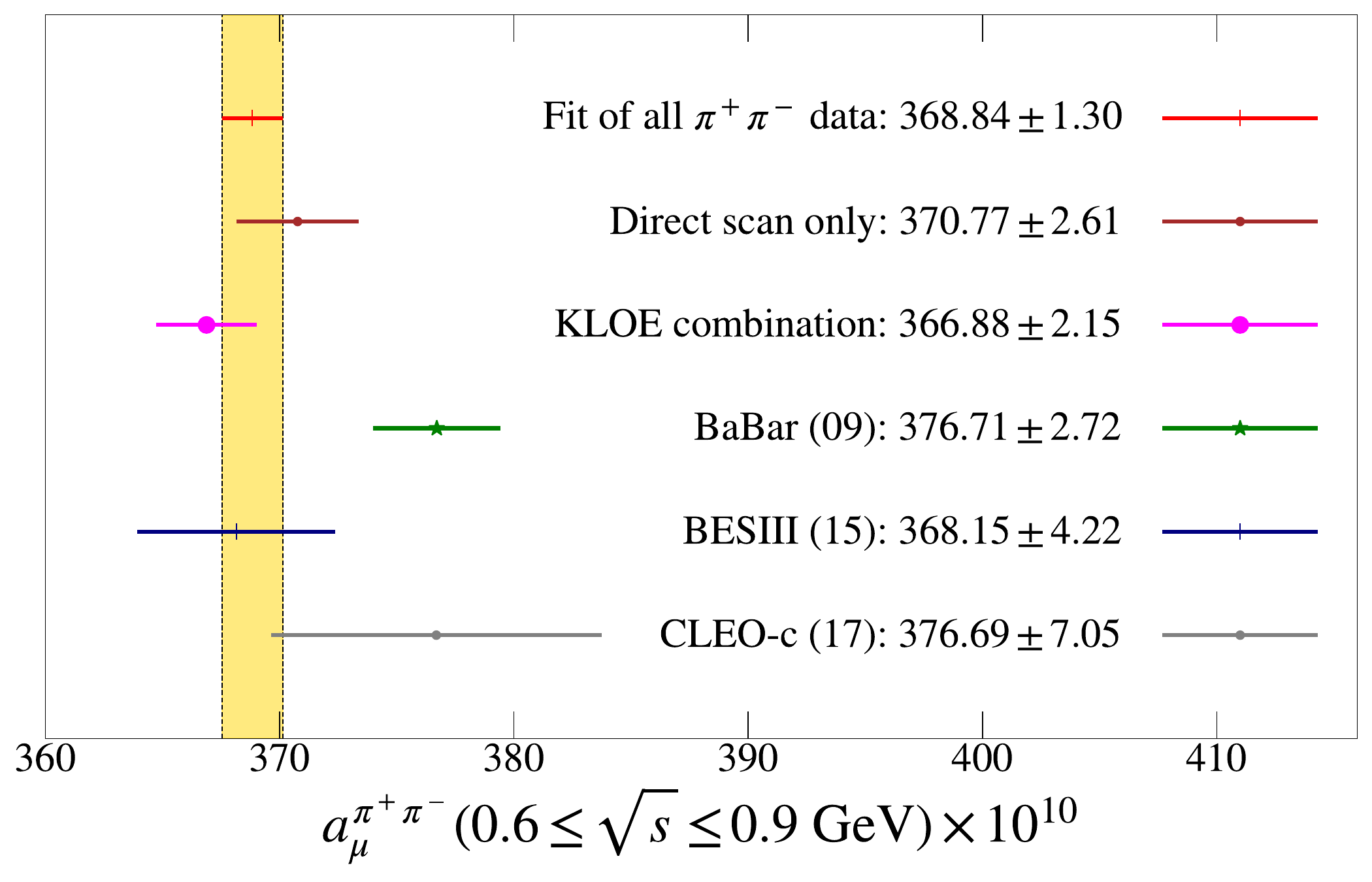}
    \hspace{0.4cm}
    \includegraphics[width=0.43\textwidth]{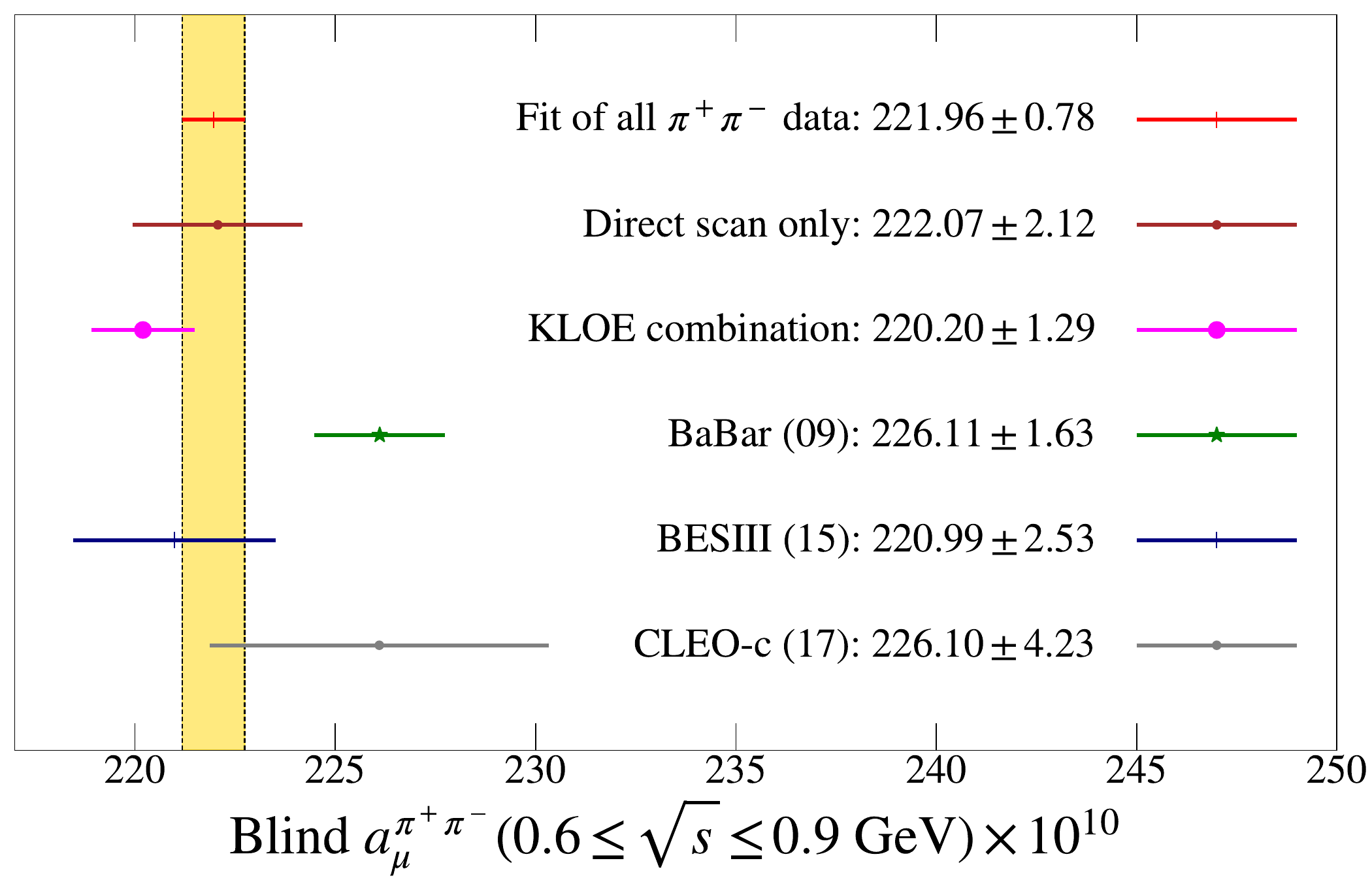}
    }\hfill 
     \caption{Comparisons of results for the $\pi^+\pi^-$ channel from KNT19~\cite{keshavarzi:2019abf} before (left) and after (right) implementing the described blinding example. Fig.~\ref{fig:pipirho_plots} is showing plotted cross section values. Fig.~\ref{fig:pipirho_ints} is comparing integrated values. As such, they are blinded with different offsets and are not comparable. Details regarding the input datasets can be found in~\cite{keshavarzi:2019abf}.} \label{fig:pipirho}
\vspace{-0.6cm}
\end{figure*}

To compare a resulting data combination with its input datasets for any derived quantity or plot, the input datasets are adjusted by the same blinding scheme only for that comparison. This allows for analysis of the combination under blinding (see e.g. Fig.~\ref{fig:pipirho_plots}). The unknown energy-dependent adjustments, which are different for derived quantities and plots, ensure that no comparison of a combination with input datasets can result in full knowledge of the influence of that dataset on a combination or consequent extraction of the blinding offsets. No input datasets are saved with blinding offsets applied.

At the level of the individual channels, the generic dispersion integral used to extract the blinded value of the HVP contribution to an observable $O_{\rm HVP}$ then has the form
\begin{align} \label{eq:int_blind}
    O^{\rm blind}_{\rm HVP} & = \int^{s_{\rm high}}_{s_{\rm low}} {\rm d}s\, f(s) \, \sigma^{\rm blind}_{\rm had}(s) \, \nonumber 
    \\
    & = a\,b\, \int^{s_{\rm high}}_{s_{\rm low}} {\rm d}s \, (s+s_0)^c \, f(s) \, \sigma_{\rm had}(s) \, ,
\end{align}
where $f(s)$ is an energy-dependent kernel function.
Observables calculated as part of the KNTW analyses other than $a_\mu^{\rm HVP}$ include (but are not restricted to) the HVP contributions to the electron and tau $g$$-$$2$, the hadronic contributions to the running electromagnetic coupling, and the HVP contributions to the hyperfine splitting of muonium (see~\cite{keshavarzi:2019abf}). Whilst the $\sigma_{\rm had}(s)$ is the same for each $O_{\rm HVP}$, the kernel functions $f(s)$ are different and induce a different energy-dependent weighting. The blinding scheme effectively adjusts this energy-dependence by a different, unknown amount for each integrated $O_{\rm HVP}$. This not only adds another layer of blinding between each observable, but also has the benefit of removing possible bias towards $a_\mu^{\rm HVP}$ being the primary output of the analysis, putting more emphasis on the combined data as the primary product and figure of merit.

A description and the allowed values for each offset are given in Table~\ref{table:offsets}. The offset $a$ is only applied to the data entering Eq.~\eqref{eq:int_blind}. The allowed values for $b$ have been chosen to not be too extreme and importantly avoid $b=1$, which would provide no overall scale factor. The values for $c$ have been chosen to moderately adjust the energy-dependence without unreasonably distorting the shape of the hadronic cross section and to avoid $c = 0$, which would provide no adjustment. The additive $s_0$ provides an additional energy-dependent shift without which it would be known that no energy-dependent blinding due to $c$ would be present at exactly $s = 1$ GeV$^2$. 

\begin{table}[t!]
\centering
\begin{tblr}{
  cells={valign=m,halign=c},
  colspec={m{2.1cm}m{2.8cm}X[c]},
  width=\linewidth,
  hlines,
  vlines
}
{\bf Offset (seed)} & {\bf Allowed values} & {\bf Comment} \\ 
$a$ ($r_1$) & $\pm 1$ & Not applied to plots \\
$b$ ($r_2$) & $0.1 \leq b \leq 0.9$ and $1.1 \leq b \leq 10$ & Avoids scaling of 1 \\
$c$ ($r_3$) & $-0.05 \leq c \leq -0.01$ and $0.01 \leq c \leq 0.05$ & Avoids power of zero \\
$s_0$ ($r_4$) & $-0.01 \leq s_0 \leq 1$ GeV$^2$ & Blinds $s^c = (1$ GeV$^2)^c$ \\
\end{tblr}
\caption{The blinding offsets applied in Eq.~\eqref{eq:sigma_blind} to the hadronic cross section data. Each offset has an associated seed used to generate a random number from the allowed values given in the second column.}
\label{table:offsets}
\vspace{-0.5cm}
\end{table}

The offsets are set by a person (blinder) external to KNTW. The blinder is provided a software package which, when executed, asks the blinder to choose (and make a private note of) five blinding seeds known only to them. Four are for the offsets listed in Table~\ref{table:offsets}. The fifth ($r_5$) results in an unknown integer in the range 1-100 that will offset the integer ID number of each hadronic channel, of which there are less than 100 in total. This ID number offset is concatenated to the beginning (for integrals) or end (for plots) of the seed for each of the other four offsets, resulting in distinct blinding for each channel and different blinding offsets for the data entering integrals and plots. Each seed can be any signed, 32-bit integer and initializes a random number generator for each offset that yields a value from the allowed ranges provided in  Table~\ref{table:offsets}. The package then automatically generates compiled and obfuscated software routines containing the offsets which the blinder provides to KNTW. In this way, neither the blinding offsets nor random seeds can be disentangled or reverse-engineered by accident or without the intention of removing the blinding entirely. 

To unblind the analysis, i.e. to remove the blinding offsets, requires correct input of the offsets on execution of the software. The blinding scheme has been devised to have two layers to allow for additional comparisons of results and systematic cross checks without fully unblinding. The unblinding will therefore happen in two stages. First, when the analyses and data combinations of all individual hadronic channels are complete, a relative unblinding will be performed where the blinder inputs only $r_5$ to the KNTW software. This removes the ID number offset for the individual channels, leaving a common blinding for all channels. In this relatively unblind stage, the final results are still concealed, but cross checks of the analysis can be performed by comparing the results from different hadronic channels under a common set of offsets. Once these final cross checks are complete, the KNTW analysis will be frozen, and no further changes will be made. Only then will the blinder be asked to input all offsets and fully unblind the analysis. Once an analysis is unblind, a new analysis (i.e. to incorporate significant changes or new hadronic cross section data) will require a blinder to repeat the process and introduce new blinding offsets to the new KNTW analysis.

%% file: sections/4_example.tex
{\bf Example implementation} - 
As an example, consider the blinder choosing the seeds $r_1 = 11111$, $r_2= 2222$, $r_3 = 333$, $r_4 = 44$ and $r_5 = 5$. The seed $r_5 = 5$ results in a value of 62 (from the range 1-100) from the random number generator. For a hadronic channel with ID number $n$, the ID number offset is then $(n+62)$. In the most recent KNT analysis (denoted KNT19)~\cite{keshavarzi:2019abf}, the $\pi^+\pi^-$ channel ID was $n=10$. In this example, the ID number offset for this channel would then be $n+62 = 72$ and the blinding seeds $\tilde{r}$ for the data entering the integrals and plots of the KNT19 $\pi^+\pi^-$ channel would be:
\begin{table}[h]
\vspace{-0.1cm}
\centering
\begin{tblr}{
  cells={valign=m,halign=c},
  colspec={X[c]X[c]X[c]X[c]X[c]},
  width=\linewidth,
  hlines,
  vlines
}
& $\tilde{r}_1$ & $\tilde{r}_2$ & $\tilde{r}_3$ & $\tilde{r}_4$ \\ 
Integrals: & 7211111 & 722222  & 72333 & 7244 \\
Plots: & - & 222272 & 33372 & 4472 \\
\end{tblr}
\end{table}
\vspace{-0.3cm} \\ These in turn result in the following randomly generated offset values:
\begin{table}[h]
\vspace{-0.1cm}
\centering
\begin{tblr}{
  cells={valign=m,halign=c},
  colspec={X[c]X[c]X[c]X[c]X[c]},
  width=\linewidth,
  hlines,
  vlines
}
& $a$ & $b$ & $c$ & $s_0$ \\ 
Integrals: & +1 & 0.617  & 0.047 & 0.013 \\
Plots: & - & 1.702 & 0.048 & 0.094 \\
\end{tblr}
\end{table}
%1 0.6168394911135217 0.046524304416999625 0.013467244748808766
%1.7021896167422939 0.04776935586852969 0.0935785529458968

Inputting these values into Eq.~\eqref{eq:sigma_blind} for the KNT19 $\pi^+\pi^-$ data combination results in Fig.~\ref{fig:pipirho}. Note that Fig.~\ref{fig:pipirho_plots} is displaying the plotted cross section and so is subject to the offset values for ``Plots". In this case, the value for multiplicative scale factor, $b=1.702$, is clearly visible, whilst the energy-dependent shifts are not discernible. Fig.~\ref{fig:pipirho_ints} is comparing integrated values and has therefore been blinded by the ``Integrals" offsets. In this case, $a=+1$ so the integrated values are positive and the value $b=0.617$ is also apparent. Here, the energy-dependent changes from $c$ and $s_0$ are noticeable in the comparisons between the input data sets and the resulting combination, which are subtly different before and after blinding. Under such a blinding scheme, the impact from changes to the analysis or the introduction of new datasets would not be fully evident until after unblinding.

\begin{figure}[t!]
    \centering
   \includegraphics[width=0.48\textwidth]{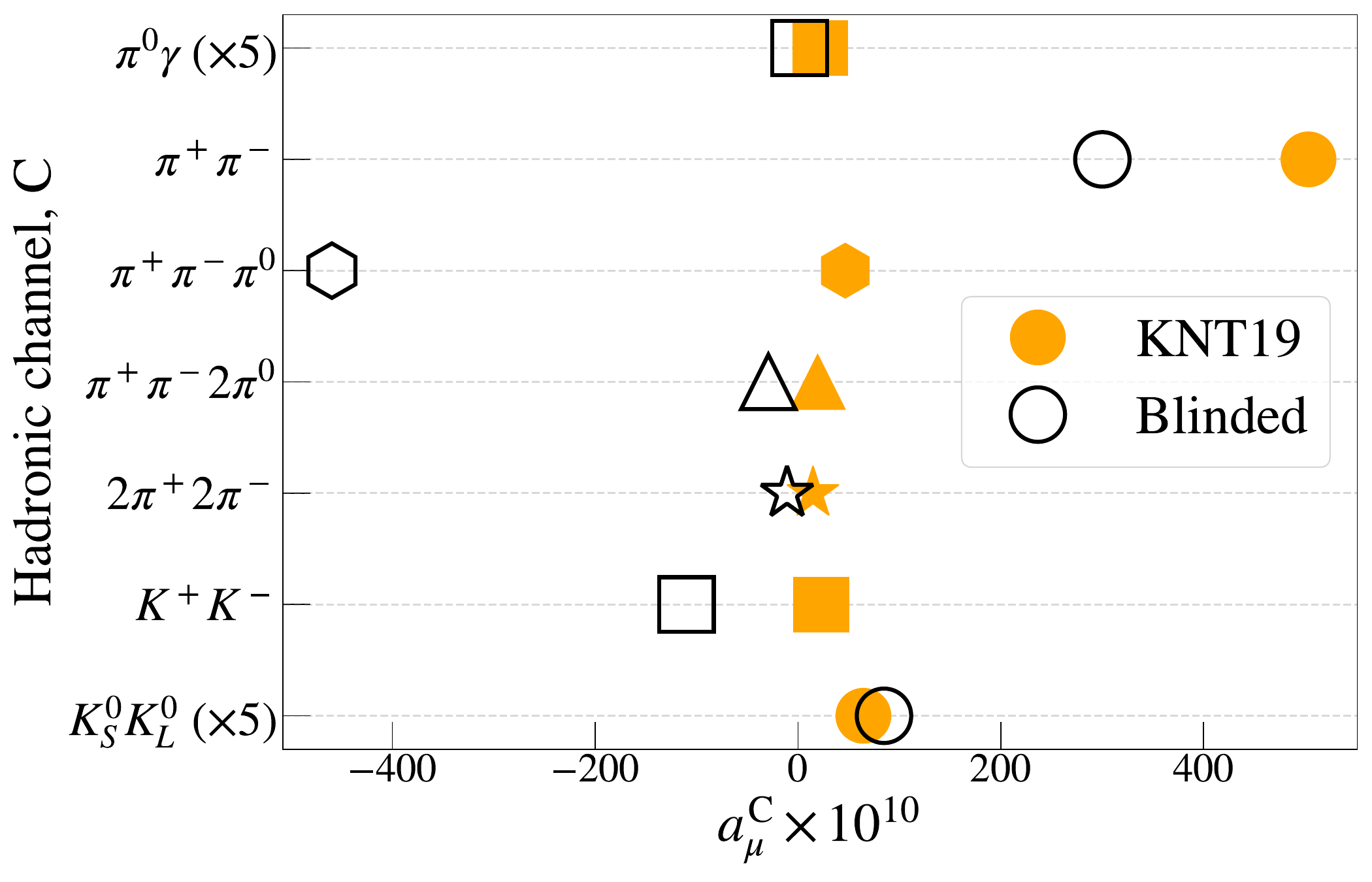}
   \caption{Values for $a_\mu^{\rm HVP}$ for different hadronic channels from the KNT19 analysis~\cite{keshavarzi:2019abf} before (filled) and after (no fill) implementing the described blinding example. In this specific example, the blind results for $\pi^0\gamma$ and $K^0_SK^0_L$ are randomly similar to their KNT19 values and have therefore been scaled by $\times5$ to make the differences visible.}
   \label{fig:blind_all}
\vspace{-0.6cm}
\end{figure}
There are $>50$ hadronic channels in the KNTW analysis each with an ID number $n$ that, when offset by an unknown integer from the seed $r_5$, results in a different set of offsets for each channel that cannot be disentangled. Continuing with the same example, the integrated values for $a_\mu^{\rm HVP}$ from the dominant hadronic channels in the KNT19 analysis are shown in Fig.~\ref{fig:blind_all}. Here, the distinct blinding for each channel is clearly evident, with some even having negative values due to the random assignment of $a=-1$. As intended, extracting knowledge of the blinding from comparing results for different channels is impossible.

%% file: sections/5_conclusions.tex
{\bf Conclusions and outlook} - 
Current tensions in different evaluations of $a_\mu^{\rm HVP}$ indicate either a discovery of new physics or a multi-method confirmation of the SM when comparing the resulting contrasting values for $a_\mu^{\rm SM}$ with $a_\mu^{\rm exp}$. Under particular scrutiny are the data-driven evaluations of $a_\mu^{\rm HVP}$~\cite{colangelo:2018mtw,hoferichter:2019mqg,davier:2019can,keshavarzi:2019abf}, which combine measured $e^+e^-\to{\rm hadrons}$ cross section data to input into dispersion integrals, allowing for the extraction of $a_\mu^{\rm HVP}$ and other observables sensitive to hadronic vacuum polarization effects. The most severe tension in the data-driven determinations is due to a measurement of the dominant $e^+e^-\to\pi^+\pi^-$ cross section by the CMD-3 experiment~\cite{CMD-3:2023alj,CMD-3:2023rfe} that has a cross section several standard deviations larger than all other previous data. Used in isolation to calculate $a_\mu^{\rm HVP}$, it results in a value for $a_\mu^{\rm SM}$ that is consistent with $a_\mu^{\rm exp}$. Analysis choices by different groups of how to combine these data lead to different results and in previous cases have been shown to differ at the level of the uncertainty on the combined cross section~\cite{Aoyama:2020ynm}.  The impact on future data-driven determinations of $a_\mu^{\rm HVP}$ from including the CMD-3 data and new, more precise experimental data (which could either resolve the current data tensions or make them worse with increased precision) will be influenced by past or future analysis choices on how to combine them.

The new KNTW analysis framework is undergoing a complete overhaul and modernization aimed at providing a new state-of-the-art in data-driven evaluations of $a_\mu^{\rm HVP}$. Given the high-stakes due to the current tensions and the crossroads in the search for new physics in the muon $g$$-$$2$, future data-driven determinations of $a_\mu^{\rm HVP}$ must attempt to avoid analysis bias wherever possible. Given that results from the Muon $g$$-$$2$ Experiment at Fermilab, lattice QCD calculations, and future $e^+e^-\to{\rm hadrons}$ cross section measurements are all blind analyses, implementing analysis blinding in data-driven determinations of $a_\mu^{\rm HVP}$ is paramount. This is crucial before including new data whose impact on the resulting $a_\mu^{\rm HVP}$ will be influenced by unavoidable analysis choices. The first blinding scheme for data-driven evaluations of the HVP has been proposed here and is in place as part of the new, ongoing KNTW analysis.

With the final results from the Muon $g$$-$$2$ Experiment at Fermilab and an updated value of $a_\mu^{\rm SM}$ from the Muon $g$$-$$2$ Theory Initiative expected soon, it is a pivotal time for the study of the muon's anomalous magnetic moment. Future KNTW analyses will improve upon previous determinations, fully re-evaluate the HVP and contribute to the studies of lepton anomalies for many years to come. To ensure their integrity, unblinding can happen only when an analysis is complete and at an appropriate time with respect to the release of new or updated hadronic cross section data.

%% file: main.bbl
%merlin.mbs apsrev4-1.bst 2010-07-25 4.21a (PWD, AO, DPC) hacked
%Control: key (0)
%Control: author (72) initials jnrlst
%Control: editor formatted (1) identically to author
%Control: production of article title (-1) disabled
%Control: page (0) single
%Control: year (1) truncated
%Control: production of eprint (0) enabled
\begin{thebibliography}{90}%
\makeatletter
\providecommand \@ifxundefined [1]{%
 \@ifx{#1\undefined}
}%
\providecommand \@ifnum [1]{%
 \ifnum #1\expandafter \@firstoftwo
 \else \expandafter \@secondoftwo
 \fi
}%
\providecommand \@ifx [1]{%
 \ifx #1\expandafter \@firstoftwo
 \else \expandafter \@secondoftwo
 \fi
}%
\providecommand \natexlab [1]{#1}%
\providecommand \enquote  [1]{``#1''}%
\providecommand \bibnamefont  [1]{#1}%
\providecommand \bibfnamefont [1]{#1}%
\providecommand \citenamefont [1]{#1}%
\providecommand \href@noop [0]{\@secondoftwo}%
\providecommand \href [0]{\begingroup \@sanitize@url \@href}%
\providecommand \@href[1]{\@@startlink{#1}\@@href}%
\providecommand \@@href[1]{\endgroup#1\@@endlink}%
\providecommand \@sanitize@url [0]{\catcode `\\12\catcode `\$12\catcode `\&12\catcode `\#12\catcode `\^12\catcode `\_12\catcode `\%12\relax}%
\providecommand \@@startlink[1]{}%
\providecommand \@@endlink[0]{}%
\providecommand \url  [0]{\begingroup\@sanitize@url \@url }%
\providecommand \@url [1]{\endgroup\@href {#1}{\urlprefix }}%
\providecommand \urlprefix  [0]{URL }%
\providecommand \Eprint [0]{\href }%
\providecommand \doibase [0]{http://dx.doi.org/}%
\providecommand \selectlanguage [0]{\@gobble}%
\providecommand \bibinfo  [0]{\@secondoftwo}%
\providecommand \bibfield  [0]{\@secondoftwo}%
\providecommand \translation [1]{[#1]}%
\providecommand \BibitemOpen [0]{}%
\providecommand \bibitemStop [0]{}%
\providecommand \bibitemNoStop [0]{.\EOS\space}%
\providecommand \EOS [0]{\spacefactor3000\relax}%
\providecommand \BibitemShut  [1]{\csname bibitem#1\endcsname}%
\let\auto@bib@innerbib\@empty
%</preamble>
\bibitem [{\citenamefont {Aoyama}\ \emph {et~al.}(2020)\citenamefont {Aoyama} \emph {et~al.}}]{Aoyama:2020ynm}%
  \BibitemOpen
  \bibfield  {author} {\bibinfo {author} {\bibfnamefont {T.}~\bibnamefont {Aoyama}} \emph {et~al.},\ }\href {\doibase 10.1016/j.physrep.2020.07.006} {\bibfield  {journal} {\bibinfo  {journal} {Phys. Rept.}\ }\textbf {\bibinfo {volume} {887}},\ \bibinfo {pages} {1} (\bibinfo {year} {2020})},\ \Eprint {http://arxiv.org/abs/2006.04822} {arXiv:2006.04822 [hep-ph]} \BibitemShut {NoStop}%
\bibitem [{\citenamefont {Aoyama}\ \emph {et~al.}(2012)\citenamefont {Aoyama}, \citenamefont {Hayakawa}, \citenamefont {Kinoshita},\ and\ \citenamefont {Nio}}]{aoyama:2012wk}%
  \BibitemOpen
  \bibfield  {author} {\bibinfo {author} {\bibfnamefont {T.}~\bibnamefont {Aoyama}}, \bibinfo {author} {\bibfnamefont {M.}~\bibnamefont {Hayakawa}}, \bibinfo {author} {\bibfnamefont {T.}~\bibnamefont {Kinoshita}}, \ and\ \bibinfo {author} {\bibfnamefont {M.}~\bibnamefont {Nio}},\ }\href {\doibase 10.1103/PhysRevLett.109.111808} {\bibfield  {journal} {\bibinfo  {journal} {Phys. Rev. Lett.}\ }\textbf {\bibinfo {volume} {109}},\ \bibinfo {pages} {111808} (\bibinfo {year} {2012})},\ \Eprint {http://arxiv.org/abs/1205.5370} {arXiv:1205.5370 [hep-ph]} \BibitemShut {NoStop}%
%%CITATION = ARXIV:1205.5370;%%
\bibitem [{\citenamefont {Aoyama}\ \emph {et~al.}(2019)\citenamefont {Aoyama}, \citenamefont {Kinoshita},\ and\ \citenamefont {Nio}}]{Aoyama:2019ryr}%
  \BibitemOpen
  \bibfield  {author} {\bibinfo {author} {\bibfnamefont {T.}~\bibnamefont {Aoyama}}, \bibinfo {author} {\bibfnamefont {T.}~\bibnamefont {Kinoshita}}, \ and\ \bibinfo {author} {\bibfnamefont {M.}~\bibnamefont {Nio}},\ }\href {\doibase 10.3390/atoms7010028} {\bibfield  {journal} {\bibinfo  {journal} {Atoms}\ }\textbf {\bibinfo {volume} {7}},\ \bibinfo {pages} {28} (\bibinfo {year} {2019})}\BibitemShut {NoStop}%
\bibitem [{\citenamefont {Czarnecki}\ \emph {et~al.}(2003)\citenamefont {Czarnecki}, \citenamefont {Marciano},\ and\ \citenamefont {Vainshtein}}]{czarnecki:2002nt}%
  \BibitemOpen
  \bibfield  {author} {\bibinfo {author} {\bibfnamefont {A.}~\bibnamefont {Czarnecki}}, \bibinfo {author} {\bibfnamefont {W.~J.}\ \bibnamefont {Marciano}}, \ and\ \bibinfo {author} {\bibfnamefont {A.}~\bibnamefont {Vainshtein}},\ }\href {\doibase 10.1103/PhysRevD.67.073006} {\bibfield  {journal} {\bibinfo  {journal} {Phys. Rev.}\ }\textbf {\bibinfo {volume} {D67}},\ \bibinfo {pages} {073006} (\bibinfo {year} {2003})},\ \bibinfo {note} {[Erratum: Phys. Rev. {\bf D73}, 119901 (2006)]},\ \Eprint {http://arxiv.org/abs/hep-ph/0212229} {arXiv:hep-ph/0212229 [hep-ph]} \BibitemShut {NoStop}%
%%CITATION = HEP-PH/0212229;%%
\bibitem [{\citenamefont {Gnendiger}\ \emph {et~al.}(2013)\citenamefont {Gnendiger}, \citenamefont {St{\"o}ckinger},\ and\ \citenamefont {St{\"o}ckinger-Kim}}]{gnendiger:2013pva}%
  \BibitemOpen
  \bibfield  {author} {\bibinfo {author} {\bibfnamefont {C.}~\bibnamefont {Gnendiger}}, \bibinfo {author} {\bibfnamefont {D.}~\bibnamefont {St{\"o}ckinger}}, \ and\ \bibinfo {author} {\bibfnamefont {H.}~\bibnamefont {St{\"o}ckinger-Kim}},\ }\href {\doibase 10.1103/PhysRevD.88.053005} {\bibfield  {journal} {\bibinfo  {journal} {Phys. Rev.}\ }\textbf {\bibinfo {volume} {D88}},\ \bibinfo {pages} {053005} (\bibinfo {year} {2013})},\ \Eprint {http://arxiv.org/abs/1306.5546} {arXiv:1306.5546 [hep-ph]} \BibitemShut {NoStop}%
%%CITATION = ARXIV:1306.5546;%%
\bibitem [{\citenamefont {Davier}\ \emph {et~al.}(2017)\citenamefont {Davier}, \citenamefont {Hoecker}, \citenamefont {Malaescu},\ and\ \citenamefont {Zhang}}]{davier:2017zfy}%
  \BibitemOpen
  \bibfield  {author} {\bibinfo {author} {\bibfnamefont {M.}~\bibnamefont {Davier}}, \bibinfo {author} {\bibfnamefont {A.}~\bibnamefont {Hoecker}}, \bibinfo {author} {\bibfnamefont {B.}~\bibnamefont {Malaescu}}, \ and\ \bibinfo {author} {\bibfnamefont {Z.}~\bibnamefont {Zhang}},\ }\href {\doibase 10.1140/epjc/s10052-017-5161-6} {\bibfield  {journal} {\bibinfo  {journal} {Eur. Phys. J.}\ }\textbf {\bibinfo {volume} {C77}},\ \bibinfo {pages} {827} (\bibinfo {year} {2017})},\ \Eprint {http://arxiv.org/abs/1706.09436} {arXiv:1706.09436 [hep-ph]} \BibitemShut {NoStop}%
%%CITATION = ARXIV:1706.09436;%%
\bibitem [{\citenamefont {Keshavarzi}\ \emph {et~al.}(2018)\citenamefont {Keshavarzi}, \citenamefont {Nomura},\ and\ \citenamefont {Teubner}}]{keshavarzi:2018mgv}%
  \BibitemOpen
  \bibfield  {author} {\bibinfo {author} {\bibfnamefont {A.}~\bibnamefont {Keshavarzi}}, \bibinfo {author} {\bibfnamefont {D.}~\bibnamefont {Nomura}}, \ and\ \bibinfo {author} {\bibfnamefont {T.}~\bibnamefont {Teubner}},\ }\href {\doibase 10.1103/PhysRevD.97.114025} {\bibfield  {journal} {\bibinfo  {journal} {Phys. Rev.}\ }\textbf {\bibinfo {volume} {D97}},\ \bibinfo {pages} {114025} (\bibinfo {year} {2018})},\ \Eprint {http://arxiv.org/abs/1802.02995} {arXiv:1802.02995 [hep-ph]} \BibitemShut {NoStop}%
%%CITATION = ARXIV:1802.02995;%%
\bibitem [{\citenamefont {Colangelo}\ \emph {et~al.}(2019)\citenamefont {Colangelo}, \citenamefont {Hoferichter},\ and\ \citenamefont {Stoffer}}]{colangelo:2018mtw}%
  \BibitemOpen
  \bibfield  {author} {\bibinfo {author} {\bibfnamefont {G.}~\bibnamefont {Colangelo}}, \bibinfo {author} {\bibfnamefont {M.}~\bibnamefont {Hoferichter}}, \ and\ \bibinfo {author} {\bibfnamefont {P.}~\bibnamefont {Stoffer}},\ }\href {\doibase 10.1007/JHEP02(2019)006} {\bibfield  {journal} {\bibinfo  {journal} {JHEP}\ }\textbf {\bibinfo {volume} {02}},\ \bibinfo {pages} {006} (\bibinfo {year} {2019})},\ \Eprint {http://arxiv.org/abs/1810.00007} {arXiv:1810.00007 [hep-ph]} \BibitemShut {NoStop}%
%%CITATION = ARXIV:1810.00007;%%
\bibitem [{\citenamefont {Hoferichter}\ \emph {et~al.}(2019)\citenamefont {Hoferichter}, \citenamefont {Hoid},\ and\ \citenamefont {Kubis}}]{hoferichter:2019mqg}%
  \BibitemOpen
  \bibfield  {author} {\bibinfo {author} {\bibfnamefont {M.}~\bibnamefont {Hoferichter}}, \bibinfo {author} {\bibfnamefont {B.-L.}\ \bibnamefont {Hoid}}, \ and\ \bibinfo {author} {\bibfnamefont {B.}~\bibnamefont {Kubis}},\ }\href {\doibase 10.1007/JHEP08(2019)137} {\bibfield  {journal} {\bibinfo  {journal} {JHEP}\ }\textbf {\bibinfo {volume} {08}},\ \bibinfo {pages} {137} (\bibinfo {year} {2019})},\ \Eprint {http://arxiv.org/abs/1907.01556} {arXiv:1907.01556 [hep-ph]} \BibitemShut {NoStop}%
%%CITATION = ARXIV:1907.01556;%%
\bibitem [{\citenamefont {Davier}\ \emph {et~al.}(2020)\citenamefont {Davier}, \citenamefont {Hoecker}, \citenamefont {Malaescu},\ and\ \citenamefont {Zhang}}]{davier:2019can}%
  \BibitemOpen
  \bibfield  {author} {\bibinfo {author} {\bibfnamefont {M.}~\bibnamefont {Davier}}, \bibinfo {author} {\bibfnamefont {A.}~\bibnamefont {Hoecker}}, \bibinfo {author} {\bibfnamefont {B.}~\bibnamefont {Malaescu}}, \ and\ \bibinfo {author} {\bibfnamefont {Z.}~\bibnamefont {Zhang}},\ }\href {\doibase 10.1140/epjc/s10052-020-7792-2} {\bibfield  {journal} {\bibinfo  {journal} {Eur. Phys. J.}\ }\textbf {\bibinfo {volume} {C80}},\ \bibinfo {pages} {241} (\bibinfo {year} {2020})},\ \bibinfo {note} {[Erratum: Eur. Phys. J. {\bf C80}, 410 (2020)]},\ \Eprint {http://arxiv.org/abs/1908.00921} {arXiv:1908.00921 [hep-ph]} \BibitemShut {NoStop}%
%%CITATION = ARXIV:1908.00921;%%
\bibitem [{\citenamefont {Keshavarzi}\ \emph {et~al.}(2020)\citenamefont {Keshavarzi}, \citenamefont {Nomura},\ and\ \citenamefont {Teubner}}]{keshavarzi:2019abf}%
  \BibitemOpen
  \bibfield  {author} {\bibinfo {author} {\bibfnamefont {A.}~\bibnamefont {Keshavarzi}}, \bibinfo {author} {\bibfnamefont {D.}~\bibnamefont {Nomura}}, \ and\ \bibinfo {author} {\bibfnamefont {T.}~\bibnamefont {Teubner}},\ }\href {\doibase 10.1103/PhysRevD.101.014029} {\bibfield  {journal} {\bibinfo  {journal} {Phys. Rev.}\ }\textbf {\bibinfo {volume} {D101}},\ \bibinfo {pages} {014029} (\bibinfo {year} {2020})},\ \Eprint {http://arxiv.org/abs/1911.00367} {arXiv:1911.00367 [hep-ph]} \BibitemShut {NoStop}%
%%CITATION = ARXIV:1911.00367;%%
\bibitem [{\citenamefont {Kurz}\ \emph {et~al.}(2014)\citenamefont {Kurz}, \citenamefont {Liu}, \citenamefont {Marquard},\ and\ \citenamefont {Steinhauser}}]{kurz:2014wya}%
  \BibitemOpen
  \bibfield  {author} {\bibinfo {author} {\bibfnamefont {A.}~\bibnamefont {Kurz}}, \bibinfo {author} {\bibfnamefont {T.}~\bibnamefont {Liu}}, \bibinfo {author} {\bibfnamefont {P.}~\bibnamefont {Marquard}}, \ and\ \bibinfo {author} {\bibfnamefont {M.}~\bibnamefont {Steinhauser}},\ }\href {\doibase 10.1016/j.physletb.2014.05.043} {\bibfield  {journal} {\bibinfo  {journal} {Phys. Lett.}\ }\textbf {\bibinfo {volume} {B734}},\ \bibinfo {pages} {144} (\bibinfo {year} {2014})},\ \Eprint {http://arxiv.org/abs/1403.6400} {arXiv:1403.6400 [hep-ph]} \BibitemShut {NoStop}%
%%CITATION = ARXIV:1403.6400;%%
\bibitem [{\citenamefont {Melnikov}\ and\ \citenamefont {Vainshtein}(2004)}]{melnikov:2003xd}%
  \BibitemOpen
  \bibfield  {author} {\bibinfo {author} {\bibfnamefont {K.}~\bibnamefont {Melnikov}}\ and\ \bibinfo {author} {\bibfnamefont {A.}~\bibnamefont {Vainshtein}},\ }\href {\doibase 10.1103/PhysRevD.70.113006} {\bibfield  {journal} {\bibinfo  {journal} {Phys. Rev.}\ }\textbf {\bibinfo {volume} {D70}},\ \bibinfo {pages} {113006} (\bibinfo {year} {2004})},\ \Eprint {http://arxiv.org/abs/hep-ph/0312226} {arXiv:hep-ph/0312226 [hep-ph]} \BibitemShut {NoStop}%
%%CITATION = HEP-PH/0312226;%%
\bibitem [{\citenamefont {Masjuan}\ and\ \citenamefont {S{\'a}nchez-Puertas}(2017)}]{masjuan:2017tvw}%
  \BibitemOpen
  \bibfield  {author} {\bibinfo {author} {\bibfnamefont {P.}~\bibnamefont {Masjuan}}\ and\ \bibinfo {author} {\bibfnamefont {P.}~\bibnamefont {S{\'a}nchez-Puertas}},\ }\href {\doibase 10.1103/PhysRevD.95.054026} {\bibfield  {journal} {\bibinfo  {journal} {Phys. Rev.}\ }\textbf {\bibinfo {volume} {D95}},\ \bibinfo {pages} {054026} (\bibinfo {year} {2017})},\ \Eprint {http://arxiv.org/abs/1701.05829} {arXiv:1701.05829 [hep-ph]} \BibitemShut {NoStop}%
%%CITATION = ARXIV:1701.05829;%%
\bibitem [{\citenamefont {Colangelo}\ \emph {et~al.}(2017)\citenamefont {Colangelo}, \citenamefont {Hoferichter}, \citenamefont {Procura},\ and\ \citenamefont {Stoffer}}]{Colangelo:2017fiz}%
  \BibitemOpen
  \bibfield  {author} {\bibinfo {author} {\bibfnamefont {G.}~\bibnamefont {Colangelo}}, \bibinfo {author} {\bibfnamefont {M.}~\bibnamefont {Hoferichter}}, \bibinfo {author} {\bibfnamefont {M.}~\bibnamefont {Procura}}, \ and\ \bibinfo {author} {\bibfnamefont {P.}~\bibnamefont {Stoffer}},\ }\href {\doibase 10.1007/JHEP04(2017)161} {\bibfield  {journal} {\bibinfo  {journal} {JHEP}\ }\textbf {\bibinfo {volume} {04}},\ \bibinfo {pages} {161} (\bibinfo {year} {2017})},\ \Eprint {http://arxiv.org/abs/1702.07347} {arXiv:1702.07347 [hep-ph]} \BibitemShut {NoStop}%
%%CITATION = ARXIV:1702.07347;%%
\bibitem [{\citenamefont {Hoferichter}\ \emph {et~al.}(2018)\citenamefont {Hoferichter}, \citenamefont {Hoid}, \citenamefont {Kubis}, \citenamefont {Leupold},\ and\ \citenamefont {Schneider}}]{hoferichter:2018kwz}%
  \BibitemOpen
  \bibfield  {author} {\bibinfo {author} {\bibfnamefont {M.}~\bibnamefont {Hoferichter}}, \bibinfo {author} {\bibfnamefont {B.-L.}\ \bibnamefont {Hoid}}, \bibinfo {author} {\bibfnamefont {B.}~\bibnamefont {Kubis}}, \bibinfo {author} {\bibfnamefont {S.}~\bibnamefont {Leupold}}, \ and\ \bibinfo {author} {\bibfnamefont {S.~P.}\ \bibnamefont {Schneider}},\ }\href {\doibase 10.1007/JHEP10(2018)141} {\bibfield  {journal} {\bibinfo  {journal} {JHEP}\ }\textbf {\bibinfo {volume} {10}},\ \bibinfo {pages} {141} (\bibinfo {year} {2018})},\ \Eprint {http://arxiv.org/abs/1808.04823} {arXiv:1808.04823 [hep-ph]} \BibitemShut {NoStop}%
%%CITATION = ARXIV:1808.04823;%%
\bibitem [{\citenamefont {G{\'e}rardin}\ \emph {et~al.}(2019)\citenamefont {G{\'e}rardin}, \citenamefont {Meyer},\ and\ \citenamefont {Nyffeler}}]{gerardin:2019vio}%
  \BibitemOpen
  \bibfield  {author} {\bibinfo {author} {\bibfnamefont {A.}~\bibnamefont {G{\'e}rardin}}, \bibinfo {author} {\bibfnamefont {H.~B.}\ \bibnamefont {Meyer}}, \ and\ \bibinfo {author} {\bibfnamefont {A.}~\bibnamefont {Nyffeler}},\ }\href {\doibase 10.1103/PhysRevD.100.034520} {\bibfield  {journal} {\bibinfo  {journal} {Phys. Rev.}\ }\textbf {\bibinfo {volume} {D100}},\ \bibinfo {pages} {034520} (\bibinfo {year} {2019})},\ \Eprint {http://arxiv.org/abs/1903.09471} {arXiv:1903.09471 [hep-lat]} \BibitemShut {NoStop}%
%%CITATION = ARXIV:1903.09471;%%
\bibitem [{\citenamefont {Bijnens}\ \emph {et~al.}(2019)\citenamefont {Bijnens}, \citenamefont {Hermansson-Truedsson},\ and\ \citenamefont {Rodr{\'i}guez-S{\'a}nchez}}]{bijnens:2019ghy}%
  \BibitemOpen
  \bibfield  {author} {\bibinfo {author} {\bibfnamefont {J.}~\bibnamefont {Bijnens}}, \bibinfo {author} {\bibfnamefont {N.}~\bibnamefont {Hermansson-Truedsson}}, \ and\ \bibinfo {author} {\bibfnamefont {A.}~\bibnamefont {Rodr{\'i}guez-S{\'a}nchez}},\ }\href {\doibase 10.1016/j.physletb.2019.134994} {\bibfield  {journal} {\bibinfo  {journal} {Phys. Lett.}\ }\textbf {\bibinfo {volume} {B798}},\ \bibinfo {pages} {134994} (\bibinfo {year} {2019})},\ \Eprint {http://arxiv.org/abs/1908.03331} {arXiv:1908.03331 [hep-ph]} \BibitemShut {NoStop}%
%%CITATION = ARXIV:1908.03331;%%
\bibitem [{\citenamefont {Colangelo}\ \emph {et~al.}(2020)\citenamefont {Colangelo}, \citenamefont {Hagelstein}, \citenamefont {Hoferichter}, \citenamefont {Laub},\ and\ \citenamefont {Stoffer}}]{colangelo:2019uex}%
  \BibitemOpen
  \bibfield  {author} {\bibinfo {author} {\bibfnamefont {G.}~\bibnamefont {Colangelo}}, \bibinfo {author} {\bibfnamefont {F.}~\bibnamefont {Hagelstein}}, \bibinfo {author} {\bibfnamefont {M.}~\bibnamefont {Hoferichter}}, \bibinfo {author} {\bibfnamefont {L.}~\bibnamefont {Laub}}, \ and\ \bibinfo {author} {\bibfnamefont {P.}~\bibnamefont {Stoffer}},\ }\href {\doibase 10.1007/JHEP03(2020)101} {\bibfield  {journal} {\bibinfo  {journal} {JHEP}\ }\textbf {\bibinfo {volume} {03}},\ \bibinfo {pages} {101} (\bibinfo {year} {2020})},\ \Eprint {http://arxiv.org/abs/1910.13432} {arXiv:1910.13432 [hep-ph]} \BibitemShut {NoStop}%
%%CITATION = ARXIV:1910.13432;%%
\bibitem [{\citenamefont {Blum}\ \emph {et~al.}(2020)\citenamefont {Blum}, \citenamefont {Christ}, \citenamefont {Hayakawa}, \citenamefont {Izubuchi}, \citenamefont {Jin}, \citenamefont {Jung},\ and\ \citenamefont {Lehner}}]{Blum:2019ugy}%
  \BibitemOpen
  \bibfield  {author} {\bibinfo {author} {\bibfnamefont {T.}~\bibnamefont {Blum}}, \bibinfo {author} {\bibfnamefont {N.}~\bibnamefont {Christ}}, \bibinfo {author} {\bibfnamefont {M.}~\bibnamefont {Hayakawa}}, \bibinfo {author} {\bibfnamefont {T.}~\bibnamefont {Izubuchi}}, \bibinfo {author} {\bibfnamefont {L.}~\bibnamefont {Jin}}, \bibinfo {author} {\bibfnamefont {C.}~\bibnamefont {Jung}}, \ and\ \bibinfo {author} {\bibfnamefont {C.}~\bibnamefont {Lehner}},\ }\href {\doibase 10.1103/PhysRevLett.124.132002} {\bibfield  {journal} {\bibinfo  {journal} {Phys. Rev. Lett.}\ }\textbf {\bibinfo {volume} {124}},\ \bibinfo {pages} {132002} (\bibinfo {year} {2020})},\ \Eprint {http://arxiv.org/abs/1911.08123} {arXiv:1911.08123 [hep-lat]} \BibitemShut {NoStop}%
%%CITATION = ARXIV:1911.08123;%%
\bibitem [{\citenamefont {Colangelo}\ \emph {et~al.}(2014)\citenamefont {Colangelo}, \citenamefont {Hoferichter}, \citenamefont {Nyffeler}, \citenamefont {Passera},\ and\ \citenamefont {Stoffer}}]{colangelo:2014qya}%
  \BibitemOpen
  \bibfield  {author} {\bibinfo {author} {\bibfnamefont {G.}~\bibnamefont {Colangelo}}, \bibinfo {author} {\bibfnamefont {M.}~\bibnamefont {Hoferichter}}, \bibinfo {author} {\bibfnamefont {A.}~\bibnamefont {Nyffeler}}, \bibinfo {author} {\bibfnamefont {M.}~\bibnamefont {Passera}}, \ and\ \bibinfo {author} {\bibfnamefont {P.}~\bibnamefont {Stoffer}},\ }\href {\doibase 10.1016/j.physletb.2014.06.012} {\bibfield  {journal} {\bibinfo  {journal} {Phys. Lett.}\ }\textbf {\bibinfo {volume} {B735}},\ \bibinfo {pages} {90} (\bibinfo {year} {2014})},\ \Eprint {http://arxiv.org/abs/1403.7512} {arXiv:1403.7512 [hep-ph]} \BibitemShut {NoStop}%
%%CITATION = ARXIV:1403.7512;%%
\bibitem [{\citenamefont {Brodsky}\ and\ \citenamefont {De~Rafael}(1968)}]{Brodsky:1967sr}%
  \BibitemOpen
  \bibfield  {author} {\bibinfo {author} {\bibfnamefont {S.~J.}\ \bibnamefont {Brodsky}}\ and\ \bibinfo {author} {\bibfnamefont {E.}~\bibnamefont {De~Rafael}},\ }\href {\doibase 10.1103/PhysRev.168.1620} {\bibfield  {journal} {\bibinfo  {journal} {Phys. Rev.}\ }\textbf {\bibinfo {volume} {168}},\ \bibinfo {pages} {1620} (\bibinfo {year} {1968})}\BibitemShut {NoStop}%
\bibitem [{\citenamefont {Lautrup}\ and\ \citenamefont {De~Rafael}(1968)}]{Lautrup:1968tdb}%
  \BibitemOpen
  \bibfield  {author} {\bibinfo {author} {\bibfnamefont {B.~E.}\ \bibnamefont {Lautrup}}\ and\ \bibinfo {author} {\bibfnamefont {E.}~\bibnamefont {De~Rafael}},\ }\href {\doibase 10.1103/PhysRev.174.1835} {\bibfield  {journal} {\bibinfo  {journal} {Phys. Rev.}\ }\textbf {\bibinfo {volume} {174}},\ \bibinfo {pages} {1835} (\bibinfo {year} {1968})}\BibitemShut {NoStop}%
\bibitem [{\citenamefont {Krause}(1997)}]{Krause:1996rf}%
  \BibitemOpen
  \bibfield  {author} {\bibinfo {author} {\bibfnamefont {B.}~\bibnamefont {Krause}},\ }\href {\doibase 10.1016/S0370-2693(96)01346-9} {\bibfield  {journal} {\bibinfo  {journal} {Phys. Lett. B}\ }\textbf {\bibinfo {volume} {390}},\ \bibinfo {pages} {392} (\bibinfo {year} {1997})},\ \Eprint {http://arxiv.org/abs/hep-ph/9607259} {arXiv:hep-ph/9607259} \BibitemShut {NoStop}%
\bibitem [{\citenamefont {Jegerlehner}(2017)}]{Jegerlehner:2017gek}%
  \BibitemOpen
  \bibfield  {author} {\bibinfo {author} {\bibfnamefont {F.}~\bibnamefont {Jegerlehner}},\ }\href {\doibase 10.1007/978-3-319-63577-4} {}Vol.\ \bibinfo {volume} {274}\ (\bibinfo  {publisher} {Springer},\ \bibinfo {address} {Cham},\ \bibinfo {year} {2017})\BibitemShut {NoStop}%
\bibitem [{\citenamefont {Jegerlehner}(2016)}]{Jegerlehner:2015stw}%
  \BibitemOpen
  \bibfield  {author} {\bibinfo {author} {\bibfnamefont {F.}~\bibnamefont {Jegerlehner}},\ }\href {\doibase 10.1051/epjconf/201611801016} {\bibfield  {journal} {\bibinfo  {journal} {EPJ Web Conf.}\ }\textbf {\bibinfo {volume} {118}},\ \bibinfo {pages} {01016} (\bibinfo {year} {2016})},\ \Eprint {http://arxiv.org/abs/1511.04473} {arXiv:1511.04473 [hep-ph]} \BibitemShut {NoStop}%
\bibitem [{\citenamefont {Jegerlehner}(2018)}]{Jegerlehner:2017lbd}%
  \BibitemOpen
  \bibfield  {author} {\bibinfo {author} {\bibfnamefont {F.}~\bibnamefont {Jegerlehner}},\ }\href {\doibase 10.1051/epjconf/201816600022} {\bibfield  {journal} {\bibinfo  {journal} {EPJ Web Conf.}\ }\textbf {\bibinfo {volume} {166}},\ \bibinfo {pages} {00022} (\bibinfo {year} {2018})},\ \Eprint {http://arxiv.org/abs/1705.00263} {arXiv:1705.00263 [hep-ph]} \BibitemShut {NoStop}%
\bibitem [{\citenamefont {Jegerlehner}(2019{\natexlab{a}})}]{Jegerlehner:2017zsb}%
  \BibitemOpen
  \bibfield  {author} {\bibinfo {author} {\bibfnamefont {F.}~\bibnamefont {Jegerlehner}},\ }\href {\doibase 10.1051/epjconf/201921801003} {\bibfield  {journal} {\bibinfo  {journal} {EPJ Web Conf.}\ }\textbf {\bibinfo {volume} {218}},\ \bibinfo {pages} {01003} (\bibinfo {year} {2019}{\natexlab{a}})},\ \Eprint {http://arxiv.org/abs/1711.06089} {arXiv:1711.06089 [hep-ph]} \BibitemShut {NoStop}%
\bibitem [{\citenamefont {Jegerlehner}(2019{\natexlab{b}})}]{Jegerlehner:2018gjd}%
  \BibitemOpen
  \bibfield  {author} {\bibinfo {author} {\bibfnamefont {F.}~\bibnamefont {Jegerlehner}},\ }\href {\doibase 10.1051/epjconf/201919901010} {\bibfield  {journal} {\bibinfo  {journal} {EPJ Web Conf.}\ }\textbf {\bibinfo {volume} {199}},\ \bibinfo {pages} {01010} (\bibinfo {year} {2019}{\natexlab{b}})},\ \Eprint {http://arxiv.org/abs/1809.07413} {arXiv:1809.07413 [hep-ph]} \BibitemShut {NoStop}%
\bibitem [{\citenamefont {Eidelman}\ and\ \citenamefont {Jegerlehner}(1995)}]{Eidelman:1995ny}%
  \BibitemOpen
  \bibfield  {author} {\bibinfo {author} {\bibfnamefont {S.}~\bibnamefont {Eidelman}}\ and\ \bibinfo {author} {\bibfnamefont {F.}~\bibnamefont {Jegerlehner}},\ }\href {\doibase 10.1007/BF01553984} {\bibfield  {journal} {\bibinfo  {journal} {Z. Phys. C}\ }\textbf {\bibinfo {volume} {67}},\ \bibinfo {pages} {585} (\bibinfo {year} {1995})},\ \Eprint {http://arxiv.org/abs/hep-ph/9502298} {arXiv:hep-ph/9502298} \BibitemShut {NoStop}%
\bibitem [{\citenamefont {Benayoun}\ \emph {et~al.}(2008)\citenamefont {Benayoun}, \citenamefont {David}, \citenamefont {DelBuono}, \citenamefont {Leitner},\ and\ \citenamefont {O'Connell}}]{Benayoun:2007cu}%
  \BibitemOpen
  \bibfield  {author} {\bibinfo {author} {\bibfnamefont {M.}~\bibnamefont {Benayoun}}, \bibinfo {author} {\bibfnamefont {P.}~\bibnamefont {David}}, \bibinfo {author} {\bibfnamefont {L.}~\bibnamefont {DelBuono}}, \bibinfo {author} {\bibfnamefont {O.}~\bibnamefont {Leitner}}, \ and\ \bibinfo {author} {\bibfnamefont {H.~B.}\ \bibnamefont {O'Connell}},\ }\href {\doibase 10.1140/epjc/s10052-008-0586-6} {\bibfield  {journal} {\bibinfo  {journal} {Eur. Phys. J. C}\ }\textbf {\bibinfo {volume} {55}},\ \bibinfo {pages} {199} (\bibinfo {year} {2008})},\ \Eprint {http://arxiv.org/abs/0711.4482} {arXiv:0711.4482 [hep-ph]} \BibitemShut {NoStop}%
\bibitem [{\citenamefont {Benayoun}\ \emph {et~al.}(2012)\citenamefont {Benayoun}, \citenamefont {David}, \citenamefont {DelBuono},\ and\ \citenamefont {Jegerlehner}}]{Benayoun:2012etq}%
  \BibitemOpen
  \bibfield  {author} {\bibinfo {author} {\bibfnamefont {M.}~\bibnamefont {Benayoun}}, \bibinfo {author} {\bibfnamefont {P.}~\bibnamefont {David}}, \bibinfo {author} {\bibfnamefont {L.}~\bibnamefont {DelBuono}}, \ and\ \bibinfo {author} {\bibfnamefont {F.}~\bibnamefont {Jegerlehner}},\ }\href {\doibase 10.1140/epjc/s10052-011-1848-2} {\bibfield  {journal} {\bibinfo  {journal} {Eur. Phys. J. C}\ }\textbf {\bibinfo {volume} {72}},\ \bibinfo {pages} {1848} (\bibinfo {year} {2012})},\ \Eprint {http://arxiv.org/abs/1106.1315} {arXiv:1106.1315 [hep-ph]} \BibitemShut {NoStop}%
\bibitem [{\citenamefont {Benayoun}\ \emph {et~al.}(2013)\citenamefont {Benayoun}, \citenamefont {David}, \citenamefont {DelBuono},\ and\ \citenamefont {Jegerlehner}}]{Benayoun:2012wc}%
  \BibitemOpen
  \bibfield  {author} {\bibinfo {author} {\bibfnamefont {M.}~\bibnamefont {Benayoun}}, \bibinfo {author} {\bibfnamefont {P.}~\bibnamefont {David}}, \bibinfo {author} {\bibfnamefont {L.}~\bibnamefont {DelBuono}}, \ and\ \bibinfo {author} {\bibfnamefont {F.}~\bibnamefont {Jegerlehner}},\ }\href {\doibase 10.1140/epjc/s10052-013-2453-3} {\bibfield  {journal} {\bibinfo  {journal} {Eur. Phys. J. C}\ }\textbf {\bibinfo {volume} {73}},\ \bibinfo {pages} {2453} (\bibinfo {year} {2013})},\ \Eprint {http://arxiv.org/abs/1210.7184} {arXiv:1210.7184 [hep-ph]} \BibitemShut {NoStop}%
\bibitem [{\citenamefont {Benayoun}\ \emph {et~al.}(2015)\citenamefont {Benayoun}, \citenamefont {David}, \citenamefont {DelBuono},\ and\ \citenamefont {Jegerlehner}}]{Benayoun:2015gxa}%
  \BibitemOpen
  \bibfield  {author} {\bibinfo {author} {\bibfnamefont {M.}~\bibnamefont {Benayoun}}, \bibinfo {author} {\bibfnamefont {P.}~\bibnamefont {David}}, \bibinfo {author} {\bibfnamefont {L.}~\bibnamefont {DelBuono}}, \ and\ \bibinfo {author} {\bibfnamefont {F.}~\bibnamefont {Jegerlehner}},\ }\href {\doibase 10.1140/epjc/s10052-015-3830-x} {\bibfield  {journal} {\bibinfo  {journal} {Eur. Phys. J. C}\ }\textbf {\bibinfo {volume} {75}},\ \bibinfo {pages} {613} (\bibinfo {year} {2015})},\ \Eprint {http://arxiv.org/abs/1507.02943} {arXiv:1507.02943 [hep-ph]} \BibitemShut {NoStop}%
\bibitem [{\citenamefont {Benayoun}\ \emph {et~al.}(2020)\citenamefont {Benayoun}, \citenamefont {Delbuono},\ and\ \citenamefont {Jegerlehner}}]{Benayoun:2019zwh}%
  \BibitemOpen
  \bibfield  {author} {\bibinfo {author} {\bibfnamefont {M.}~\bibnamefont {Benayoun}}, \bibinfo {author} {\bibfnamefont {L.}~\bibnamefont {Delbuono}}, \ and\ \bibinfo {author} {\bibfnamefont {F.}~\bibnamefont {Jegerlehner}},\ }\href {\doibase 10.1140/epjc/s10052-020-7611-9} {\bibfield  {journal} {\bibinfo  {journal} {Eur. Phys. J. C}\ }\textbf {\bibinfo {volume} {80}},\ \bibinfo {pages} {81} (\bibinfo {year} {2020})},\ \bibinfo {note} {[Erratum: Eur.Phys.J.C 80, 244 (2020)]},\ \Eprint {http://arxiv.org/abs/1903.11034} {arXiv:1903.11034 [hep-ph]} \BibitemShut {NoStop}%
\bibitem [{\citenamefont {Davier}\ \emph {et~al.}(2011)\citenamefont {Davier}, \citenamefont {Hoecker}, \citenamefont {Malaescu},\ and\ \citenamefont {Zhang}}]{Davier:2010nc}%
  \BibitemOpen
  \bibfield  {author} {\bibinfo {author} {\bibfnamefont {M.}~\bibnamefont {Davier}}, \bibinfo {author} {\bibfnamefont {A.}~\bibnamefont {Hoecker}}, \bibinfo {author} {\bibfnamefont {B.}~\bibnamefont {Malaescu}}, \ and\ \bibinfo {author} {\bibfnamefont {Z.}~\bibnamefont {Zhang}},\ }\href {\doibase 10.1140/epjc/s10052-012-1874-8} {\bibfield  {journal} {\bibinfo  {journal} {Eur. Phys. J. C}\ }\textbf {\bibinfo {volume} {71}},\ \bibinfo {pages} {1515} (\bibinfo {year} {2011})},\ \bibinfo {note} {[Erratum: Eur.Phys.J.C 72, 1874 (2012)]},\ \Eprint {http://arxiv.org/abs/1010.4180} {arXiv:1010.4180 [hep-ph]} \BibitemShut {NoStop}%
\bibitem [{\citenamefont {Hagiwara}\ \emph {et~al.}(2004)\citenamefont {Hagiwara}, \citenamefont {Martin}, \citenamefont {Nomura},\ and\ \citenamefont {Teubner}}]{Hagiwara:2003da}%
  \BibitemOpen
  \bibfield  {author} {\bibinfo {author} {\bibfnamefont {K.}~\bibnamefont {Hagiwara}}, \bibinfo {author} {\bibfnamefont {A.~D.}\ \bibnamefont {Martin}}, \bibinfo {author} {\bibfnamefont {D.}~\bibnamefont {Nomura}}, \ and\ \bibinfo {author} {\bibfnamefont {T.}~\bibnamefont {Teubner}},\ }\href {\doibase 10.1103/PhysRevD.69.093003} {\bibfield  {journal} {\bibinfo  {journal} {Phys. Rev. D}\ }\textbf {\bibinfo {volume} {69}},\ \bibinfo {pages} {093003} (\bibinfo {year} {2004})},\ \Eprint {http://arxiv.org/abs/hep-ph/0312250} {arXiv:hep-ph/0312250} \BibitemShut {NoStop}%
\bibitem [{\citenamefont {Hagiwara}\ \emph {et~al.}(2007)\citenamefont {Hagiwara}, \citenamefont {Martin}, \citenamefont {Nomura},\ and\ \citenamefont {Teubner}}]{Hagiwara:2006jt}%
  \BibitemOpen
  \bibfield  {author} {\bibinfo {author} {\bibfnamefont {K.}~\bibnamefont {Hagiwara}}, \bibinfo {author} {\bibfnamefont {A.~D.}\ \bibnamefont {Martin}}, \bibinfo {author} {\bibfnamefont {D.}~\bibnamefont {Nomura}}, \ and\ \bibinfo {author} {\bibfnamefont {T.}~\bibnamefont {Teubner}},\ }\href {\doibase 10.1016/j.physletb.2007.04.012} {\bibfield  {journal} {\bibinfo  {journal} {Phys. Lett. B}\ }\textbf {\bibinfo {volume} {649}},\ \bibinfo {pages} {173} (\bibinfo {year} {2007})},\ \Eprint {http://arxiv.org/abs/hep-ph/0611102} {arXiv:hep-ph/0611102} \BibitemShut {NoStop}%
\bibitem [{\citenamefont {Hagiwara}\ \emph {et~al.}(2011)\citenamefont {Hagiwara}, \citenamefont {Liao}, \citenamefont {Martin}, \citenamefont {Nomura},\ and\ \citenamefont {Teubner}}]{Hagiwara:2011af}%
  \BibitemOpen
  \bibfield  {author} {\bibinfo {author} {\bibfnamefont {K.}~\bibnamefont {Hagiwara}}, \bibinfo {author} {\bibfnamefont {R.}~\bibnamefont {Liao}}, \bibinfo {author} {\bibfnamefont {A.~D.}\ \bibnamefont {Martin}}, \bibinfo {author} {\bibfnamefont {D.}~\bibnamefont {Nomura}}, \ and\ \bibinfo {author} {\bibfnamefont {T.}~\bibnamefont {Teubner}},\ }\href {\doibase 10.1088/0954-3899/38/8/085003} {\bibfield  {journal} {\bibinfo  {journal} {J. Phys. G}\ }\textbf {\bibinfo {volume} {38}},\ \bibinfo {pages} {085003} (\bibinfo {year} {2011})},\ \Eprint {http://arxiv.org/abs/1105.3149} {arXiv:1105.3149 [hep-ph]} \BibitemShut {NoStop}%
\bibitem [{\citenamefont {Aguillard}\ \emph {et~al.}(2023)\citenamefont {Aguillard} \emph {et~al.}}]{Muong-2:2023cdq}%
  \BibitemOpen
  \bibfield  {author} {\bibinfo {author} {\bibfnamefont {D.~P.}\ \bibnamefont {Aguillard}} \emph {et~al.} (\bibinfo {collaboration} {Muon g-2}),\ }\href {\doibase 10.1103/PhysRevLett.131.161802} {\bibfield  {journal} {\bibinfo  {journal} {Phys. Rev. Lett.}\ }\textbf {\bibinfo {volume} {131}},\ \bibinfo {pages} {161802} (\bibinfo {year} {2023})},\ \Eprint {http://arxiv.org/abs/2308.06230} {arXiv:2308.06230 [hep-ex]} \BibitemShut {NoStop}%
\bibitem [{\citenamefont {Aguillard}\ \emph {et~al.}(2024)\citenamefont {Aguillard} \emph {et~al.}}]{Muong-2:2024hpx}%
  \BibitemOpen
  \bibfield  {author} {\bibinfo {author} {\bibfnamefont {D.~P.}\ \bibnamefont {Aguillard}} \emph {et~al.} (\bibinfo {collaboration} {Muon g-2}),\ }\href {\doibase 10.1103/PhysRevD.110.032009} {\bibfield  {journal} {\bibinfo  {journal} {Phys. Rev. D}\ }\textbf {\bibinfo {volume} {110}},\ \bibinfo {pages} {032009} (\bibinfo {year} {2024})},\ \Eprint {http://arxiv.org/abs/2402.15410} {arXiv:2402.15410 [hep-ex]} \BibitemShut {NoStop}%
\bibitem [{\citenamefont {Abi}\ \emph {et~al.}(2021)\citenamefont {Abi} \emph {et~al.}}]{Muong-2:2021ojo}%
  \BibitemOpen
  \bibfield  {author} {\bibinfo {author} {\bibfnamefont {B.}~\bibnamefont {Abi}} \emph {et~al.} (\bibinfo {collaboration} {Muon g-2}),\ }\href {\doibase 10.1103/PhysRevLett.126.141801} {\bibfield  {journal} {\bibinfo  {journal} {Phys. Rev. Lett.}\ }\textbf {\bibinfo {volume} {126}},\ \bibinfo {pages} {141801} (\bibinfo {year} {2021})},\ \Eprint {http://arxiv.org/abs/2104.03281} {arXiv:2104.03281 [hep-ex]} \BibitemShut {NoStop}%
\bibitem [{\citenamefont {Albahri}\ \emph {et~al.}(2021{\natexlab{a}})\citenamefont {Albahri} \emph {et~al.}}]{Muong-2:2021xzz}%
  \BibitemOpen
  \bibfield  {author} {\bibinfo {author} {\bibfnamefont {T.}~\bibnamefont {Albahri}} \emph {et~al.} (\bibinfo {collaboration} {Muon g-2}),\ }\href {\doibase 10.1103/PhysRevAccelBeams.24.044002} {\bibfield  {journal} {\bibinfo  {journal} {Phys. Rev. Accel. Beams}\ }\textbf {\bibinfo {volume} {24}},\ \bibinfo {pages} {044002} (\bibinfo {year} {2021}{\natexlab{a}})},\ \Eprint {http://arxiv.org/abs/2104.03240} {arXiv:2104.03240 [physics.acc-ph]} \BibitemShut {NoStop}%
\bibitem [{\citenamefont {Albahri}\ \emph {et~al.}(2021{\natexlab{b}})\citenamefont {Albahri} \emph {et~al.}}]{Muong-2:2021ovs}%
  \BibitemOpen
  \bibfield  {author} {\bibinfo {author} {\bibfnamefont {T.}~\bibnamefont {Albahri}} \emph {et~al.} (\bibinfo {collaboration} {Muon g-2}),\ }\href {\doibase 10.1103/PhysRevA.103.042208} {\bibfield  {journal} {\bibinfo  {journal} {Phys. Rev. A}\ }\textbf {\bibinfo {volume} {103}},\ \bibinfo {pages} {042208} (\bibinfo {year} {2021}{\natexlab{b}})},\ \Eprint {http://arxiv.org/abs/2104.03201} {arXiv:2104.03201 [hep-ex]} \BibitemShut {NoStop}%
\bibitem [{\citenamefont {Albahri}\ \emph {et~al.}(2021{\natexlab{c}})\citenamefont {Albahri} \emph {et~al.}}]{Muong-2:2021vma}%
  \BibitemOpen
  \bibfield  {author} {\bibinfo {author} {\bibfnamefont {T.}~\bibnamefont {Albahri}} \emph {et~al.} (\bibinfo {collaboration} {Muon g-2}),\ }\href {\doibase 10.1103/PhysRevD.103.072002} {\bibfield  {journal} {\bibinfo  {journal} {Phys. Rev. D}\ }\textbf {\bibinfo {volume} {103}},\ \bibinfo {pages} {072002} (\bibinfo {year} {2021}{\natexlab{c}})},\ \Eprint {http://arxiv.org/abs/2104.03247} {arXiv:2104.03247 [hep-ex]} \BibitemShut {NoStop}%
\bibitem [{\citenamefont {Bennett}\ \emph {et~al.}(2006)\citenamefont {Bennett} \emph {et~al.}}]{Muong-2:2006rrc}%
  \BibitemOpen
  \bibfield  {author} {\bibinfo {author} {\bibfnamefont {G.~W.}\ \bibnamefont {Bennett}} \emph {et~al.} (\bibinfo {collaboration} {Muon g-2}),\ }\href {\doibase 10.1103/PhysRevD.73.072003} {\bibfield  {journal} {\bibinfo  {journal} {Phys. Rev. D}\ }\textbf {\bibinfo {volume} {73}},\ \bibinfo {pages} {072003} (\bibinfo {year} {2006})},\ \Eprint {http://arxiv.org/abs/hep-ex/0602035} {arXiv:hep-ex/0602035} \BibitemShut {NoStop}%
\bibitem [{\citenamefont {Bennett}\ \emph {et~al.}(2002)\citenamefont {Bennett} \emph {et~al.}}]{Muong-2:2002wip}%
  \BibitemOpen
  \bibfield  {author} {\bibinfo {author} {\bibfnamefont {G.~W.}\ \bibnamefont {Bennett}} \emph {et~al.} (\bibinfo {collaboration} {Muon g-2}),\ }\href {\doibase 10.1103/PhysRevLett.89.101804} {\bibfield  {journal} {\bibinfo  {journal} {Phys. Rev. Lett.}\ }\textbf {\bibinfo {volume} {89}},\ \bibinfo {pages} {101804} (\bibinfo {year} {2002})},\ \bibinfo {note} {[Erratum: Phys.Rev.Lett. 89, 129903 (2002)]},\ \Eprint {http://arxiv.org/abs/hep-ex/0208001} {arXiv:hep-ex/0208001} \BibitemShut {NoStop}%
\bibitem [{\citenamefont {Bennett}\ \emph {et~al.}(2004)\citenamefont {Bennett} \emph {et~al.}}]{Muong-2:2004fok}%
  \BibitemOpen
  \bibfield  {author} {\bibinfo {author} {\bibfnamefont {G.~W.}\ \bibnamefont {Bennett}} \emph {et~al.} (\bibinfo {collaboration} {Muon g-2}),\ }\href {\doibase 10.1103/PhysRevLett.92.161802} {\bibfield  {journal} {\bibinfo  {journal} {Phys. Rev. Lett.}\ }\textbf {\bibinfo {volume} {92}},\ \bibinfo {pages} {161802} (\bibinfo {year} {2004})},\ \Eprint {http://arxiv.org/abs/hep-ex/0401008} {arXiv:hep-ex/0401008} \BibitemShut {NoStop}%
\bibitem [{\citenamefont {Abe}\ \emph {et~al.}(2019)\citenamefont {Abe} \emph {et~al.}}]{Abe:2019thb}%
  \BibitemOpen
  \bibfield  {author} {\bibinfo {author} {\bibfnamefont {M.}~\bibnamefont {Abe}} \emph {et~al.},\ }\href {\doibase 10.1093/ptep/ptz030} {\bibfield  {journal} {\bibinfo  {journal} {PTEP}\ }\textbf {\bibinfo {volume} {2019}},\ \bibinfo {pages} {053C02} (\bibinfo {year} {2019})},\ \Eprint {http://arxiv.org/abs/1901.03047} {arXiv:1901.03047 [physics.ins-det]} \BibitemShut {NoStop}%
\bibitem [{\citenamefont {Adelmann}\ \emph {et~al.}(2021)\citenamefont {Adelmann} \emph {et~al.}}]{Adelmann:2021udj}%
  \BibitemOpen
  \bibfield  {author} {\bibinfo {author} {\bibfnamefont {A.}~\bibnamefont {Adelmann}} \emph {et~al.},\ }\href@noop {} {\  (\bibinfo {year} {2021})},\ \Eprint {http://arxiv.org/abs/2102.08838} {arXiv:2102.08838 [hep-ex]} \BibitemShut {NoStop}%
\bibitem [{\citenamefont {Borsanyi}\ \emph {et~al.}(2018)\citenamefont {Borsanyi} \emph {et~al.}}]{Budapest-Marseille-Wuppertal:2017okr}%
  \BibitemOpen
  \bibfield  {author} {\bibinfo {author} {\bibfnamefont {S.}~\bibnamefont {Borsanyi}} \emph {et~al.} (\bibinfo {collaboration} {Budapest-Marseille-Wuppertal}),\ }\href {\doibase 10.1103/PhysRevLett.121.022002} {\bibfield  {journal} {\bibinfo  {journal} {Phys. Rev. Lett.}\ }\textbf {\bibinfo {volume} {121}},\ \bibinfo {pages} {022002} (\bibinfo {year} {2018})},\ \Eprint {http://arxiv.org/abs/1711.04980} {arXiv:1711.04980 [hep-lat]} \BibitemShut {NoStop}%
\bibitem [{\citenamefont {Blum}\ \emph {et~al.}(2018{\natexlab{a}})\citenamefont {Blum}, \citenamefont {Boyle}, \citenamefont {G\"ulpers}, \citenamefont {Izubuchi}, \citenamefont {Jin}, \citenamefont {Jung}, \citenamefont {J\"uttner}, \citenamefont {Lehner}, \citenamefont {Portelli},\ and\ \citenamefont {Tsang}}]{RBC:2018dos}%
  \BibitemOpen
  \bibfield  {author} {\bibinfo {author} {\bibfnamefont {T.}~\bibnamefont {Blum}}, \bibinfo {author} {\bibfnamefont {P.~A.}\ \bibnamefont {Boyle}}, \bibinfo {author} {\bibfnamefont {V.}~\bibnamefont {G\"ulpers}}, \bibinfo {author} {\bibfnamefont {T.}~\bibnamefont {Izubuchi}}, \bibinfo {author} {\bibfnamefont {L.}~\bibnamefont {Jin}}, \bibinfo {author} {\bibfnamefont {C.}~\bibnamefont {Jung}}, \bibinfo {author} {\bibfnamefont {A.}~\bibnamefont {J\"uttner}}, \bibinfo {author} {\bibfnamefont {C.}~\bibnamefont {Lehner}}, \bibinfo {author} {\bibfnamefont {A.}~\bibnamefont {Portelli}}, \ and\ \bibinfo {author} {\bibfnamefont {J.~T.}\ \bibnamefont {Tsang}} (\bibinfo {collaboration} {RBC, UKQCD}),\ }\href {\doibase 10.1103/PhysRevLett.121.022003} {\bibfield  {journal} {\bibinfo  {journal} {Phys. Rev. Lett.}\ }\textbf {\bibinfo {volume} {121}},\ \bibinfo {pages} {022003} (\bibinfo {year} {2018}{\natexlab{a}})},\ \Eprint {http://arxiv.org/abs/1801.07224} {arXiv:1801.07224 [hep-lat]} \BibitemShut {NoStop}%
\bibitem [{\citenamefont {Giusti}\ \emph {et~al.}(2019)\citenamefont {Giusti}, \citenamefont {Lubicz}, \citenamefont {Martinelli}, \citenamefont {Sanfilippo},\ and\ \citenamefont {Simula}}]{Giusti:2019xct}%
  \BibitemOpen
  \bibfield  {author} {\bibinfo {author} {\bibfnamefont {D.}~\bibnamefont {Giusti}}, \bibinfo {author} {\bibfnamefont {V.}~\bibnamefont {Lubicz}}, \bibinfo {author} {\bibfnamefont {G.}~\bibnamefont {Martinelli}}, \bibinfo {author} {\bibfnamefont {F.}~\bibnamefont {Sanfilippo}}, \ and\ \bibinfo {author} {\bibfnamefont {S.}~\bibnamefont {Simula}},\ }\href {\doibase 10.1103/PhysRevD.99.114502} {\bibfield  {journal} {\bibinfo  {journal} {Phys. Rev. D}\ }\textbf {\bibinfo {volume} {99}},\ \bibinfo {pages} {114502} (\bibinfo {year} {2019})},\ \Eprint {http://arxiv.org/abs/1901.10462} {arXiv:1901.10462 [hep-lat]} \BibitemShut {NoStop}%
\bibitem [{\citenamefont {Davies}\ \emph {et~al.}(2020)\citenamefont {Davies} \emph {et~al.}}]{FermilabLattice:2019ugu}%
  \BibitemOpen
  \bibfield  {author} {\bibinfo {author} {\bibfnamefont {C.~T.~H.}\ \bibnamefont {Davies}} \emph {et~al.} (\bibinfo {collaboration} {Fermilab Lattice, LATTICE-HPQCD, MILC}),\ }\href {\doibase 10.1103/PhysRevD.101.034512} {\bibfield  {journal} {\bibinfo  {journal} {Phys. Rev. D}\ }\textbf {\bibinfo {volume} {101}},\ \bibinfo {pages} {034512} (\bibinfo {year} {2020})},\ \Eprint {http://arxiv.org/abs/1902.04223} {arXiv:1902.04223 [hep-lat]} \BibitemShut {NoStop}%
\bibitem [{\citenamefont {G\'erardin}\ \emph {et~al.}(2019)\citenamefont {G\'erardin}, \citenamefont {C\`e}, \citenamefont {von Hippel}, \citenamefont {H\"orz}, \citenamefont {Meyer}, \citenamefont {Mohler}, \citenamefont {Ottnad}, \citenamefont {Wilhelm},\ and\ \citenamefont {Wittig}}]{Gerardin:2019rua}%
  \BibitemOpen
  \bibfield  {author} {\bibinfo {author} {\bibfnamefont {A.}~\bibnamefont {G\'erardin}}, \bibinfo {author} {\bibfnamefont {M.}~\bibnamefont {C\`e}}, \bibinfo {author} {\bibfnamefont {G.}~\bibnamefont {von Hippel}}, \bibinfo {author} {\bibfnamefont {B.}~\bibnamefont {H\"orz}}, \bibinfo {author} {\bibfnamefont {H.~B.}\ \bibnamefont {Meyer}}, \bibinfo {author} {\bibfnamefont {D.}~\bibnamefont {Mohler}}, \bibinfo {author} {\bibfnamefont {K.}~\bibnamefont {Ottnad}}, \bibinfo {author} {\bibfnamefont {J.}~\bibnamefont {Wilhelm}}, \ and\ \bibinfo {author} {\bibfnamefont {H.}~\bibnamefont {Wittig}},\ }\href {\doibase 10.1103/PhysRevD.100.014510} {\bibfield  {journal} {\bibinfo  {journal} {Phys. Rev. D}\ }\textbf {\bibinfo {volume} {100}},\ \bibinfo {pages} {014510} (\bibinfo {year} {2019})},\ \Eprint {http://arxiv.org/abs/1904.03120} {arXiv:1904.03120 [hep-lat]} \BibitemShut {NoStop}%
\bibitem [{\citenamefont {Chakraborty}\ \emph {et~al.}(2018)\citenamefont {Chakraborty} \emph {et~al.}}]{chakraborty:2017tqp}%
  \BibitemOpen
  \bibfield  {author} {\bibinfo {author} {\bibfnamefont {B.}~\bibnamefont {Chakraborty}} \emph {et~al.} (\bibinfo {collaboration} {Fermilab Lattice, LATTICE-HPQCD, MILC}),\ }\href {\doibase 10.1103/PhysRevLett.120.152001} {\bibfield  {journal} {\bibinfo  {journal} {Phys. Rev. Lett.}\ }\textbf {\bibinfo {volume} {120}},\ \bibinfo {pages} {152001} (\bibinfo {year} {2018})},\ \Eprint {http://arxiv.org/abs/1710.11212} {arXiv:1710.11212 [hep-lat]} \BibitemShut {NoStop}%
%%CITATION = ARXIV:1710.11212;%%
\bibitem [{\citenamefont {Blum}\ \emph {et~al.}(2018{\natexlab{b}})\citenamefont {Blum}, \citenamefont {Boyle}, \citenamefont {G{\"u}lpers}, \citenamefont {Izubuchi}, \citenamefont {Jin}, \citenamefont {Jung}, \citenamefont {J{\"u}ttner}, \citenamefont {Lehner}, \citenamefont {Portelli},\ and\ \citenamefont {Tsang}}]{blum:2018mom}%
  \BibitemOpen
  \bibfield  {author} {\bibinfo {author} {\bibfnamefont {T.}~\bibnamefont {Blum}}, \bibinfo {author} {\bibfnamefont {P.~A.}\ \bibnamefont {Boyle}}, \bibinfo {author} {\bibfnamefont {V.}~\bibnamefont {G{\"u}lpers}}, \bibinfo {author} {\bibfnamefont {T.}~\bibnamefont {Izubuchi}}, \bibinfo {author} {\bibfnamefont {L.}~\bibnamefont {Jin}}, \bibinfo {author} {\bibfnamefont {C.}~\bibnamefont {Jung}}, \bibinfo {author} {\bibfnamefont {A.}~\bibnamefont {J{\"u}ttner}}, \bibinfo {author} {\bibfnamefont {C.}~\bibnamefont {Lehner}}, \bibinfo {author} {\bibfnamefont {A.}~\bibnamefont {Portelli}}, \ and\ \bibinfo {author} {\bibfnamefont {J.~T.}\ \bibnamefont {Tsang}} (\bibinfo {collaboration} {RBC, UKQCD}),\ }\href {\doibase 10.1103/PhysRevLett.121.022003} {\bibfield  {journal} {\bibinfo  {journal} {Phys. Rev. Lett.}\ }\textbf {\bibinfo {volume} {121}},\ \bibinfo {pages} {022003} (\bibinfo {year} {2018}{\natexlab{b}})},\ \Eprint {http://arxiv.org/abs/1801.07224} {arXiv:1801.07224 [hep-lat]} \BibitemShut {NoStop}%
%%CITATION = ARXIV:1801.07224;%%
\bibitem [{\citenamefont {Shintani}\ and\ \citenamefont {Kuramashi}(2019)}]{shintani:2019wai}%
  \BibitemOpen
  \bibfield  {author} {\bibinfo {author} {\bibfnamefont {E.}~\bibnamefont {Shintani}}\ and\ \bibinfo {author} {\bibfnamefont {Y.}~\bibnamefont {Kuramashi}},\ }\href {\doibase 10.1103/PhysRevD.100.034517} {\bibfield  {journal} {\bibinfo  {journal} {Phys. Rev.}\ }\textbf {\bibinfo {volume} {D100}},\ \bibinfo {pages} {034517} (\bibinfo {year} {2019})},\ \Eprint {http://arxiv.org/abs/1902.00885} {arXiv:1902.00885 [hep-lat]} \BibitemShut {NoStop}%
%%CITATION = ARXIV:1902.00885;%%
\bibitem [{\citenamefont {Aubin}\ \emph {et~al.}(2020)\citenamefont {Aubin}, \citenamefont {Blum}, \citenamefont {Tu}, \citenamefont {Golterman}, \citenamefont {Jung},\ and\ \citenamefont {Peris}}]{Aubin:2019usy}%
  \BibitemOpen
  \bibfield  {author} {\bibinfo {author} {\bibfnamefont {C.}~\bibnamefont {Aubin}}, \bibinfo {author} {\bibfnamefont {T.}~\bibnamefont {Blum}}, \bibinfo {author} {\bibfnamefont {C.}~\bibnamefont {Tu}}, \bibinfo {author} {\bibfnamefont {M.}~\bibnamefont {Golterman}}, \bibinfo {author} {\bibfnamefont {C.}~\bibnamefont {Jung}}, \ and\ \bibinfo {author} {\bibfnamefont {S.}~\bibnamefont {Peris}},\ }\href {\doibase 10.1103/PhysRevD.101.014503} {\bibfield  {journal} {\bibinfo  {journal} {Phys. Rev.}\ }\textbf {\bibinfo {volume} {D101}},\ \bibinfo {pages} {014503} (\bibinfo {year} {2020})},\ \Eprint {http://arxiv.org/abs/1905.09307} {arXiv:1905.09307 [hep-lat]} \BibitemShut {NoStop}%
%%CITATION = ARXIV:1905.09307;%%
\bibitem [{\citenamefont {Giusti}\ and\ \citenamefont {Simula}(2019)}]{giusti:2019hkz}%
  \BibitemOpen
  \bibfield  {author} {\bibinfo {author} {\bibfnamefont {D.}~\bibnamefont {Giusti}}\ and\ \bibinfo {author} {\bibfnamefont {S.}~\bibnamefont {Simula}},\ }\href {\doibase 10.22323/1.363.0104} {\bibfield  {journal} {\bibinfo  {journal} {PoS}\ }\textbf {\bibinfo {volume} {LATTICE2019}},\ \bibinfo {pages} {104} (\bibinfo {year} {2019})},\ \Eprint {http://arxiv.org/abs/1910.03874} {arXiv:1910.03874 [hep-lat]} \BibitemShut {NoStop}%
%%CITATION = ARXIV:1910.03874;%%
\bibitem [{\citenamefont {Borsanyi}\ \emph {et~al.}(2021)\citenamefont {Borsanyi} \emph {et~al.}}]{Borsanyi:2020mff}%
  \BibitemOpen
  \bibfield  {author} {\bibinfo {author} {\bibfnamefont {S.}~\bibnamefont {Borsanyi}} \emph {et~al.},\ }\href {\doibase 10.1038/s41586-021-03418-1} {\bibfield  {journal} {\bibinfo  {journal} {Nature}\ }\textbf {\bibinfo {volume} {593}},\ \bibinfo {pages} {51} (\bibinfo {year} {2021})},\ \Eprint {http://arxiv.org/abs/2002.12347} {arXiv:2002.12347 [hep-lat]} \BibitemShut {NoStop}%
\bibitem [{\citenamefont {Boccaletti}\ \emph {et~al.}(2024)\citenamefont {Boccaletti} \emph {et~al.}}]{Boccaletti:2024guq}%
  \BibitemOpen
  \bibfield  {author} {\bibinfo {author} {\bibfnamefont {A.}~\bibnamefont {Boccaletti}} \emph {et~al.},\ }\href@noop {} {\  (\bibinfo {year} {2024})},\ \Eprint {http://arxiv.org/abs/2407.10913} {arXiv:2407.10913 [hep-lat]} \BibitemShut {NoStop}%
\bibitem [{\citenamefont {Lehner}\ and\ \citenamefont {Meyer}(2020)}]{Lehner:2020crt}%
  \BibitemOpen
  \bibfield  {author} {\bibinfo {author} {\bibfnamefont {C.}~\bibnamefont {Lehner}}\ and\ \bibinfo {author} {\bibfnamefont {A.~S.}\ \bibnamefont {Meyer}},\ }\href {\doibase 10.1103/PhysRevD.101.074515} {\bibfield  {journal} {\bibinfo  {journal} {Phys. Rev. D}\ }\textbf {\bibinfo {volume} {101}},\ \bibinfo {pages} {074515} (\bibinfo {year} {2020})},\ \Eprint {http://arxiv.org/abs/2003.04177} {arXiv:2003.04177 [hep-lat]} \BibitemShut {NoStop}%
\bibitem [{\citenamefont {Alexandrou}\ \emph {et~al.}(2023)\citenamefont {Alexandrou} \emph {et~al.}}]{ExtendedTwistedMass:2022jpw}%
  \BibitemOpen
  \bibfield  {author} {\bibinfo {author} {\bibfnamefont {C.}~\bibnamefont {Alexandrou}} \emph {et~al.} (\bibinfo {collaboration} {Extended Twisted Mass}),\ }\href {\doibase 10.1103/PhysRevD.107.074506} {\bibfield  {journal} {\bibinfo  {journal} {Phys. Rev. D}\ }\textbf {\bibinfo {volume} {107}},\ \bibinfo {pages} {074506} (\bibinfo {year} {2023})},\ \Eprint {http://arxiv.org/abs/2206.15084} {arXiv:2206.15084 [hep-lat]} \BibitemShut {NoStop}%
\bibitem [{\citenamefont {Blum}\ \emph {et~al.}(2023)\citenamefont {Blum} \emph {et~al.}}]{RBC:2023pvn}%
  \BibitemOpen
  \bibfield  {author} {\bibinfo {author} {\bibfnamefont {T.}~\bibnamefont {Blum}} \emph {et~al.} (\bibinfo {collaboration} {RBC, UKQCD}),\ }\href {\doibase 10.1103/PhysRevD.108.054507} {\bibfield  {journal} {\bibinfo  {journal} {Phys. Rev. D}\ }\textbf {\bibinfo {volume} {108}},\ \bibinfo {pages} {054507} (\bibinfo {year} {2023})},\ \Eprint {http://arxiv.org/abs/2301.08696} {arXiv:2301.08696 [hep-lat]} \BibitemShut {NoStop}%
\bibitem [{\citenamefont {Kuberski}\ \emph {et~al.}(2024)\citenamefont {Kuberski}, \citenamefont {C\`e}, \citenamefont {von Hippel}, \citenamefont {Meyer}, \citenamefont {Ottnad}, \citenamefont {Risch},\ and\ \citenamefont {Wittig}}]{Kuberski:2024bcj}%
  \BibitemOpen
  \bibfield  {author} {\bibinfo {author} {\bibfnamefont {S.}~\bibnamefont {Kuberski}}, \bibinfo {author} {\bibfnamefont {M.}~\bibnamefont {C\`e}}, \bibinfo {author} {\bibfnamefont {G.}~\bibnamefont {von Hippel}}, \bibinfo {author} {\bibfnamefont {H.~B.}\ \bibnamefont {Meyer}}, \bibinfo {author} {\bibfnamefont {K.}~\bibnamefont {Ottnad}}, \bibinfo {author} {\bibfnamefont {A.}~\bibnamefont {Risch}}, \ and\ \bibinfo {author} {\bibfnamefont {H.}~\bibnamefont {Wittig}},\ }\href {\doibase 10.1007/JHEP03(2024)172} {\bibfield  {journal} {\bibinfo  {journal} {JHEP}\ }\textbf {\bibinfo {volume} {03}},\ \bibinfo {pages} {172} (\bibinfo {year} {2024})},\ \Eprint {http://arxiv.org/abs/2401.11895} {arXiv:2401.11895 [hep-lat]} \BibitemShut {NoStop}%
\bibitem [{\citenamefont {Davies}\ \emph {et~al.}(2022)\citenamefont {Davies} \emph {et~al.}}]{FermilabLattice:2022izv}%
  \BibitemOpen
  \bibfield  {author} {\bibinfo {author} {\bibfnamefont {C.~T.~H.}\ \bibnamefont {Davies}} \emph {et~al.} (\bibinfo {collaboration} {Fermilab Lattice, MILC, HPQCD}),\ }\href {\doibase 10.1103/PhysRevD.106.074509} {\bibfield  {journal} {\bibinfo  {journal} {Phys. Rev. D}\ }\textbf {\bibinfo {volume} {106}},\ \bibinfo {pages} {074509} (\bibinfo {year} {2022})},\ \Eprint {http://arxiv.org/abs/2207.04765} {arXiv:2207.04765 [hep-lat]} \BibitemShut {NoStop}%
\bibitem [{\citenamefont {Ignatov}\ \emph {et~al.}(2024{\natexlab{a}})\citenamefont {Ignatov} \emph {et~al.}}]{CMD-3:2023alj}%
  \BibitemOpen
  \bibfield  {author} {\bibinfo {author} {\bibfnamefont {F.~V.}\ \bibnamefont {Ignatov}} \emph {et~al.} (\bibinfo {collaboration} {CMD-3}),\ }\href {\doibase 10.1103/PhysRevD.109.112002} {\bibfield  {journal} {\bibinfo  {journal} {Phys. Rev. D}\ }\textbf {\bibinfo {volume} {109}},\ \bibinfo {pages} {112002} (\bibinfo {year} {2024}{\natexlab{a}})},\ \Eprint {http://arxiv.org/abs/2302.08834} {arXiv:2302.08834 [hep-ex]} \BibitemShut {NoStop}%
\bibitem [{\citenamefont {Ignatov}\ \emph {et~al.}(2024{\natexlab{b}})\citenamefont {Ignatov} \emph {et~al.}}]{CMD-3:2023rfe}%
  \BibitemOpen
  \bibfield  {author} {\bibinfo {author} {\bibfnamefont {F.~V.}\ \bibnamefont {Ignatov}} \emph {et~al.} (\bibinfo {collaboration} {CMD-3}),\ }\href {\doibase 10.1103/PhysRevLett.132.231903} {\bibfield  {journal} {\bibinfo  {journal} {Phys. Rev. Lett.}\ }\textbf {\bibinfo {volume} {132}},\ \bibinfo {pages} {231903} (\bibinfo {year} {2024}{\natexlab{b}})},\ \Eprint {http://arxiv.org/abs/2309.12910} {arXiv:2309.12910 [hep-ex]} \BibitemShut {NoStop}%
\bibitem [{\citenamefont {Carloni~Calame}\ \emph {et~al.}(2015)\citenamefont {Carloni~Calame}, \citenamefont {Passera}, \citenamefont {Trentadue},\ and\ \citenamefont {Venanzoni}}]{CarloniCalame:2015obs}%
  \BibitemOpen
  \bibfield  {author} {\bibinfo {author} {\bibfnamefont {C.~M.}\ \bibnamefont {Carloni~Calame}}, \bibinfo {author} {\bibfnamefont {M.}~\bibnamefont {Passera}}, \bibinfo {author} {\bibfnamefont {L.}~\bibnamefont {Trentadue}}, \ and\ \bibinfo {author} {\bibfnamefont {G.}~\bibnamefont {Venanzoni}},\ }\href {\doibase 10.1016/j.physletb.2015.05.020} {\bibfield  {journal} {\bibinfo  {journal} {Phys. Lett. B}\ }\textbf {\bibinfo {volume} {746}},\ \bibinfo {pages} {325} (\bibinfo {year} {2015})},\ \Eprint {http://arxiv.org/abs/1504.02228} {arXiv:1504.02228 [hep-ph]} \BibitemShut {NoStop}%
\bibitem [{\citenamefont {Abbiendi}\ \emph {et~al.}(2017)\citenamefont {Abbiendi} \emph {et~al.}}]{Abbiendi:2016xup}%
  \BibitemOpen
  \bibfield  {author} {\bibinfo {author} {\bibfnamefont {G.}~\bibnamefont {Abbiendi}} \emph {et~al.},\ }\href {\doibase 10.1140/epjc/s10052-017-4633-z} {\bibfield  {journal} {\bibinfo  {journal} {Eur. Phys. J. C}\ }\textbf {\bibinfo {volume} {77}},\ \bibinfo {pages} {139} (\bibinfo {year} {2017})},\ \Eprint {http://arxiv.org/abs/1609.08987} {arXiv:1609.08987 [hep-ex]} \BibitemShut {NoStop}%
\bibitem [{\citenamefont {Abbiendi}(2019)}]{Abbiendi:2677471}%
  \BibitemOpen
  \bibfield  {author} {\bibinfo {author} {\bibfnamefont {G.}~\bibnamefont {Abbiendi}},\ }\href {https://cds.cern.ch/record/2677471} {\emph {\bibinfo {title} {{Letter of Intent: the MUonE project}}}},\ \bibinfo {type} {Tech. Rep.}\ (\bibinfo  {institution} {CERN},\ \bibinfo {address} {Geneva},\ \bibinfo {year} {2019})\BibitemShut {NoStop}%
\bibitem [{\citenamefont {{STRONG-2020}}(2023)}]{Strong2020}%
  \BibitemOpen
  \bibfield  {author} {\bibinfo {author} {\bibnamefont {{STRONG-2020}}},\ }\href@noop {} {\enquote {\bibinfo {title} {5th {W}ork{S}top/{T}hink{S}tart: {R}adiative corrections and {M}onte {C}arlo tools for {STRONG}-2020},}\ }\bibinfo {howpublished} {2023 {M}eeting, \textsc{url:}~\url{https://indico.psi.ch/event/13707/}} (\bibinfo {year} {2023})\BibitemShut {NoStop}%
\bibitem [{\citenamefont {{R}adio{M}onte{Ca}rlow 2}(2024)}]{RadioMonteCarlow}%
  \BibitemOpen
  \bibfield  {author} {\bibinfo {author} {\bibnamefont {{R}adio{M}onte{Ca}rlow 2}},\ }\href@noop {} {\enquote {\bibinfo {title} {The {R}adio{M}onte{Ca}rlow 2 {E}ffort},}\ }\bibinfo {howpublished} {\textsc{url:}~\url{https://radiomontecarlow2.gitlab.io/}} (\bibinfo {year} {2024})\BibitemShut {NoStop}%
\bibitem [{\citenamefont {Lees}\ \emph {et~al.}(2023)\citenamefont {Lees} \emph {et~al.}}]{BaBar:2023xiy}%
  \BibitemOpen
  \bibfield  {author} {\bibinfo {author} {\bibfnamefont {J.~P.}\ \bibnamefont {Lees}} \emph {et~al.} (\bibinfo {collaboration} {BaBar}),\ }\href {\doibase 10.1103/PhysRevD.108.L111103} {\bibfield  {journal} {\bibinfo  {journal} {Phys. Rev. D}\ }\textbf {\bibinfo {volume} {108}},\ \bibinfo {pages} {L111103} (\bibinfo {year} {2023})},\ \Eprint {http://arxiv.org/abs/2308.05233} {arXiv:2308.05233 [hep-ex]} \BibitemShut {NoStop}%
\bibitem [{\citenamefont {Davier}\ \emph {et~al.}(2024)\citenamefont {Davier}, \citenamefont {Hoecker}, \citenamefont {Lutz}, \citenamefont {Malaescu},\ and\ \citenamefont {Zhang}}]{Davier:2023fpl}%
  \BibitemOpen
  \bibfield  {author} {\bibinfo {author} {\bibfnamefont {M.}~\bibnamefont {Davier}}, \bibinfo {author} {\bibfnamefont {A.}~\bibnamefont {Hoecker}}, \bibinfo {author} {\bibfnamefont {A.-M.}\ \bibnamefont {Lutz}}, \bibinfo {author} {\bibfnamefont {B.}~\bibnamefont {Malaescu}}, \ and\ \bibinfo {author} {\bibfnamefont {Z.}~\bibnamefont {Zhang}},\ }\href {\doibase 10.1140/epjc/s10052-024-12964-7} {\bibfield  {journal} {\bibinfo  {journal} {Eur. Phys. J. C}\ }\textbf {\bibinfo {volume} {84}},\ \bibinfo {pages} {721} (\bibinfo {year} {2024})},\ \Eprint {http://arxiv.org/abs/2312.02053} {arXiv:2312.02053 [hep-ph]} \BibitemShut {NoStop}%
\bibitem [{\citenamefont {{Zhiqing ZHANG}}(2024)}]{newBaBar}%
  \BibitemOpen
  \bibfield  {author} {\bibinfo {author} {\bibnamefont {{Zhiqing ZHANG}}},\ }\href@noop {} {\enquote {\bibinfo {title} {News from {BABAR}},}\ }\bibinfo {howpublished} {{M}uon $g$$-$$2$ {T}heory {I}nitiative {S}pring 2024 {M}eeting, \textsc{url:}~\url{https://indico.cern.ch/event/1400808/contributions/5902094/attachments/2842140/4968393/Zhang_HVP240422.pdf}} (\bibinfo {year} {2024})\BibitemShut {NoStop}%
\bibitem [{\citenamefont {{Hisaki Hayashii}}(2024)}]{newBelleII}%
  \BibitemOpen
  \bibfield  {author} {\bibinfo {author} {\bibnamefont {{Hisaki Hayashii}}},\ }\href@noop {} {\enquote {\bibinfo {title} {Status and plans regarding $g$$-$$2$ at belle {II}},}\ }\bibinfo {howpublished} {{M}uon $g$$-$$2$ {T}heory {I}nitiative {S}pring 2024 {M}eeting, \textsc{url:}~\url{https://indico.cern.ch/event/1400808/contributions/5902135/attachments/2841580/4968265/g-2-theory_mni-workshop_240422_v2.pdf}} (\bibinfo {year} {2024})\BibitemShut {NoStop}%
\bibitem [{\citenamefont {{Riccardo Aliberti}}(2024)}]{newBESIII}%
  \BibitemOpen
  \bibfield  {author} {\bibinfo {author} {\bibnamefont {{Riccardo Aliberti}}},\ }\href@noop {} {\enquote {\bibinfo {title} {{S}tatus and {P}lans for {E}xperimental {I}nputs to {HVP} at {BESIII}},}\ }\bibinfo {howpublished} {{M}uon $g$$-$$2$ {T}heory {I}nitiative {S}pring 2024 {M}eeting, \textsc{url:}~\url{https://indico.cern.ch/event/1400808/contributions/5902095/attachments/2841881/4967869/Muon_g-2_spring_meeting_2024.pdf}} (\bibinfo {year} {2024})\BibitemShut {NoStop}%
\bibitem [{\citenamefont {{Ivan Logashenko}}(2024)}]{newCMD3}%
  \BibitemOpen
  \bibfield  {author} {\bibinfo {author} {\bibnamefont {{Ivan Logashenko}}},\ }\href@noop {} {\enquote {\bibinfo {title} {{CMD2/3} report (on $\pi^+\pi^-$)},}\ }\bibinfo {howpublished} {{M}uon $g$$-$$2$ {T}heory {I}nitiative {S}pring 2024 {M}eeting, \textsc{url:}~\url{https://indico.cern.ch/event/1400808/contributions/5902096/attachments/2842193/4968485/Logashenko_CMD_TI_2024.pdf}} (\bibinfo {year} {2024})\BibitemShut {NoStop}%
\bibitem [{\citenamefont {{Giuseppe Mandaglio}}(2023)}]{newKLOE}%
  \BibitemOpen
  \bibfield  {author} {\bibinfo {author} {\bibnamefont {{Giuseppe Mandaglio}}},\ }\href@noop {} {\enquote {\bibinfo {title} {Hadron physics results at {KLOE-2}},}\ }\bibinfo {howpublished} {{S}ixth Plenary Workshop of the {M}uon $g$$-$$2$ {T}heory {I}nitiative (2023), \textsc{url:}~\url{https://indico.cern.ch/event/1400808/contributions/5902134/attachments/2841711/4967471/talk_IL.pdf}} (\bibinfo {year} {2023})\BibitemShut {NoStop}%
\bibitem [{\citenamefont {{Andrey Kupich}}(2024)}]{newSND}%
  \BibitemOpen
  \bibfield  {author} {\bibinfo {author} {\bibnamefont {{Andrey Kupich}}},\ }\href@noop {} {\enquote {\bibinfo {title} {Preliminary results of the $e^+e^- \to \pi^+\pi^-$ analysis with {SND} at {VEPP-2000}},}\ }\bibinfo {howpublished} {{M}uon $g$$-$$2$ {T}heory {I}nitiative {S}pring 2024 {M}eeting, \textsc{url:}~\url{https://indico.cern.ch/event/1400808/contributions/5902134/attachments/2841711/4967471/talk_IL.pdf}} (\bibinfo {year} {2024})\BibitemShut {NoStop}%
\bibitem [{\citenamefont {Ambrosino}\ \emph {et~al.}(2009)\citenamefont {Ambrosino} \emph {et~al.}}]{KLOE:2008fmq}%
  \BibitemOpen
  \bibfield  {author} {\bibinfo {author} {\bibfnamefont {F.}~\bibnamefont {Ambrosino}} \emph {et~al.} (\bibinfo {collaboration} {KLOE}),\ }\href {\doibase 10.1016/j.physletb.2008.10.060} {\bibfield  {journal} {\bibinfo  {journal} {Phys. Lett. B}\ }\textbf {\bibinfo {volume} {670}},\ \bibinfo {pages} {285} (\bibinfo {year} {2009})},\ \Eprint {http://arxiv.org/abs/0809.3950} {arXiv:0809.3950 [hep-ex]} \BibitemShut {NoStop}%
\bibitem [{\citenamefont {Ambrosino}\ \emph {et~al.}(2011)\citenamefont {Ambrosino} \emph {et~al.}}]{KLOE:2010qei}%
  \BibitemOpen
  \bibfield  {author} {\bibinfo {author} {\bibfnamefont {F.}~\bibnamefont {Ambrosino}} \emph {et~al.} (\bibinfo {collaboration} {KLOE}),\ }\href {\doibase 10.1016/j.physletb.2011.04.055} {\bibfield  {journal} {\bibinfo  {journal} {Phys. Lett. B}\ }\textbf {\bibinfo {volume} {700}},\ \bibinfo {pages} {102} (\bibinfo {year} {2011})},\ \Eprint {http://arxiv.org/abs/1006.5313} {arXiv:1006.5313 [hep-ex]} \BibitemShut {NoStop}%
\bibitem [{\citenamefont {Babusci}\ \emph {et~al.}(2013)\citenamefont {Babusci} \emph {et~al.}}]{KLOE:2012anl}%
  \BibitemOpen
  \bibfield  {author} {\bibinfo {author} {\bibfnamefont {D.}~\bibnamefont {Babusci}} \emph {et~al.} (\bibinfo {collaboration} {KLOE}),\ }\href {\doibase 10.1016/j.physletb.2013.02.029} {\bibfield  {journal} {\bibinfo  {journal} {Phys. Lett. B}\ }\textbf {\bibinfo {volume} {720}},\ \bibinfo {pages} {336} (\bibinfo {year} {2013})},\ \Eprint {http://arxiv.org/abs/1212.4524} {arXiv:1212.4524 [hep-ex]} \BibitemShut {NoStop}%
\bibitem [{\citenamefont {Anastasi}\ \emph {et~al.}(2018)\citenamefont {Anastasi} \emph {et~al.}}]{KLOE-2:2017fda}%
  \BibitemOpen
  \bibfield  {author} {\bibinfo {author} {\bibfnamefont {A.}~\bibnamefont {Anastasi}} \emph {et~al.} (\bibinfo {collaboration} {KLOE-2}),\ }\href {\doibase 10.1007/JHEP03(2018)173} {\bibfield  {journal} {\bibinfo  {journal} {JHEP}\ }\textbf {\bibinfo {volume} {03}},\ \bibinfo {pages} {173} (\bibinfo {year} {2018})},\ \Eprint {http://arxiv.org/abs/1711.03085} {arXiv:1711.03085 [hep-ex]} \BibitemShut {NoStop}%
\bibitem [{\citenamefont {Aubert}\ \emph {et~al.}(2009)\citenamefont {Aubert} \emph {et~al.}}]{BaBar:2009wpw}%
  \BibitemOpen
  \bibfield  {author} {\bibinfo {author} {\bibfnamefont {B.}~\bibnamefont {Aubert}} \emph {et~al.} (\bibinfo {collaboration} {BaBar}),\ }\href {\doibase 10.1103/PhysRevLett.103.231801} {\bibfield  {journal} {\bibinfo  {journal} {Phys. Rev. Lett.}\ }\textbf {\bibinfo {volume} {103}},\ \bibinfo {pages} {231801} (\bibinfo {year} {2009})},\ \Eprint {http://arxiv.org/abs/0908.3589} {arXiv:0908.3589 [hep-ex]} \BibitemShut {NoStop}%
\bibitem [{\citenamefont {Lees}\ \emph {et~al.}(2012)\citenamefont {Lees} \emph {et~al.}}]{BaBar:2012bdw}%
  \BibitemOpen
  \bibfield  {author} {\bibinfo {author} {\bibfnamefont {J.~P.}\ \bibnamefont {Lees}} \emph {et~al.} (\bibinfo {collaboration} {BaBar}),\ }\href {\doibase 10.1103/PhysRevD.86.032013} {\bibfield  {journal} {\bibinfo  {journal} {Phys. Rev. D}\ }\textbf {\bibinfo {volume} {86}},\ \bibinfo {pages} {032013} (\bibinfo {year} {2012})},\ \Eprint {http://arxiv.org/abs/1205.2228} {arXiv:1205.2228 [hep-ex]} \BibitemShut {NoStop}%
\bibitem [{\citenamefont {Masjuan}\ \emph {et~al.}(2024)\citenamefont {Masjuan}, \citenamefont {Miranda},\ and\ \citenamefont {Roig}}]{Masjuan:2023qsp}%
  \BibitemOpen
  \bibfield  {author} {\bibinfo {author} {\bibfnamefont {P.}~\bibnamefont {Masjuan}}, \bibinfo {author} {\bibfnamefont {A.}~\bibnamefont {Miranda}}, \ and\ \bibinfo {author} {\bibfnamefont {P.}~\bibnamefont {Roig}},\ }\href {\doibase 10.1016/j.physletb.2024.138492} {\bibfield  {journal} {\bibinfo  {journal} {Phys. Lett. B}\ }\textbf {\bibinfo {volume} {850}},\ \bibinfo {pages} {138492} (\bibinfo {year} {2024})},\ \Eprint {http://arxiv.org/abs/2305.20005} {arXiv:2305.20005 [hep-ph]} \BibitemShut {NoStop}%
\bibitem [{\citenamefont {Bruno}\ \emph {et~al.}(2018)\citenamefont {Bruno}, \citenamefont {Izubuchi}, \citenamefont {Lehner},\ and\ \citenamefont {Meyer}}]{Bruno:2018ono}%
  \BibitemOpen
  \bibfield  {author} {\bibinfo {author} {\bibfnamefont {M.}~\bibnamefont {Bruno}}, \bibinfo {author} {\bibfnamefont {T.}~\bibnamefont {Izubuchi}}, \bibinfo {author} {\bibfnamefont {C.}~\bibnamefont {Lehner}}, \ and\ \bibinfo {author} {\bibfnamefont {A.}~\bibnamefont {Meyer}},\ }\href {\doibase 10.22323/1.334.0135} {\bibfield  {journal} {\bibinfo  {journal} {PoS}\ }\textbf {\bibinfo {volume} {LATTICE2018}},\ \bibinfo {pages} {135} (\bibinfo {year} {2018})},\ \Eprint {http://arxiv.org/abs/1811.00508} {arXiv:1811.00508 [hep-lat]} \BibitemShut {NoStop}%
\end{thebibliography}%
